\documentclass[10pt,twocolumn,letterpaper]{article}

\usepackage[pagenumbers]{cvpr} % To force page numbers, e.g. for an arXiv version

\definecolor{cvprblue}{rgb}{0.21,0.49,0.74}
\usepackage[pagebackref,breaklinks,colorlinks,allcolors=cvprblue]{hyperref}
\usepackage{multirow}
\usepackage{fontawesome5}
\usepackage{arydshln}
\usepackage{microtype}
\usepackage[letterpaper,margin=0.75in]{geometry}
\usepackage{booktabs}

\def\confName{CVPR}
\def\confYear{2027}

\title{\textsc{Angel}: Anchor-Structured Neural Gaussians for Explicit Fetal Diffusion MRI Slice-to-Volume Reconstruction with Cross-Subject Learning}

\author{
Wenxuan Wu$^{1}$ \quad
Lucy Hall$^{1}$ \quad
Jonathan O'Muircheartaigh$^{1,2}$ \quad
Joseph V. Hajnal$^{1}$\\
J-Donald Tournier$^{1}$\quad
Maria Deprez$^{1}$\\ [6pt]
$^{1}$Research Department of Early Life
Imaging, School of Biomedical Engineering and \\ Imaging Sciences, King’s College London, London, UK \\
$^{2}$Institute of Psychiatry, Psychology and Neuroscience, King’s College London, London, UK\\
}

\begin{document}
\maketitle
\begin{abstract}
Diffusion magnetic resonance imaging (dMRI) provides a noninvasive view of developing tissue microstructure and white matter tracts in the fetal brain. During acquisition, however, fetal motion disrupts spatial correspondence between slices, diffusion encoding directions and signal intensities. 
% During acquisition, however, fetal motion disrupts spatial correspondence between slices and rotates diffusion-encoding directions relative to the anatomy. Recovering a coherent spatial-angular signal is further complicated by low signal-to-noise ratio (SNR) and slice-dependent intensity inhomogeneity. 
%Explicit Gaussian representations have enabled fast and accurate slice-to-volume reconstruction (SVR) of structural MRI.  Directly extending this representation to fetal dMRI lead independently optimized Gaussians to absorb measurement noise and slice-dependent bias in their angular profiles. 
We introduce ANGEL, an anchor-structured neural Gaussian framework for accurate slice-to-volume reconstruction (SVR) of fetal dMRI, that simultaneously estimates diffusion signal, fetal motion and intensity artefacts. Unlike standard explicit Gaussian representation, ANGEL is designed to prevent absorption of noise and artefacts into the reconstructed dMRI by learning a compact representation of spatio-angular signal via a shared decoder that predicts 3D Gaussians 
%carrying the dMRI signal 
across anchors and subjects. The individual anatomical variability is compactly represented through subject-specific anchors.
%that retains explicit rendering while replacing independent angular fitting with shared decoding. Each subject retains its own anchors, while decoders shared across anchors and subjects generate signal Gaussians carrying spherical harmonic (SH) coefficients. This shared parameterization help constrain higher-order angular profiles, reducing their tendency to fit local noise without requiring anatomical correspondence between subjects. A cohort-normalized latent reparameterization  further supports stable adaptation to unseen subjects. Each anchor also supports one broad, slice-conditioned bias Gaussian alongside multiple compact signal Gaussians, modeling intensity variation separately while keeping it spatially linked to the anatomy. 
Experiments on simulated neonatal and in vivo fetal data show significant quantitative improvement of dMRI reconstruction in both spatial and angular domains, in comparison to explicit Gaussian representation, implicit neural representation as well as classical methods. Our qualitative results suggest, that ANGEL supports more reliable and anatomically accurate tractograpy. 

\end{abstract}    
\section{Introduction}
\label{sec:intro}

Diffusion magnetic resonance imaging (dMRI) provides a non-invasive view of the developing fetal brain by measuring the directional diffusion of water within tissue~\cite{intro-diffusion-mri1,intro-diffusion-mri2,intro-diffusion-mri3,intro-diffusion-mri4,intro-diffusion-mri5,intro-diffusion-mri6}. However, unpredictable fetal motion during acquisition disrupts spatial
correspondence between slices, while its rotational component changes the
diffusion-encoding direction relative to the fetal anatomy~\cite{diffusion-mri-recon, diffusion-mri-recon-shard}. The acquired measurements are therefore inconsistent in both the spatial and angular domains. This is further complicated by low signal-to-noise ratio (SNR) and slice-dependent intensity inhomogeneity. Fetal dMRI slice-to-volume reconstruction (SVR) must recover a coherent spatial-angular signal from these motion- and intensity-corrupted slices to enable reliable microstructural analysis.

An established approach to is to model the underlying volume, slice motion, and intensity effects, then combine them through an acquisition model to synthesize the observed slices~\cite{classic-1,classic-2,svrtk,jo1}. Matching these predictions to the measurements provides a self-supervised physics-informed objective without requiring a ground truth motion-free reference volume. 
Classical methods optimize the reconstructed volume on a voxel grid for each acquisition separately~\cite{classic-1,classic-2,svrtk,jo1}, whereas implicit neural representations (INR) fit a separate continuous network to each subject under the same self-supervised consistency principle~\cite{nesvor,meta-inr}. For
structural MRI, the reconstructed field assigns a scalar intensity to each
spatial location. Fetal dMRI additionally requires a direction-dependent signal, commonly represented by spherical harmonics (SH)~\cite{diffusion-mri-recon, jo2, shinr}.

More recently, Gaussian primitives have enabled efficient, acquisition-aware
reconstruction of fetal structural MRI~\cite{svr-gaussian1,svr-gaussian2,svr-gaussian-fast}. Extending this representation to
dMRI would require each Gaussian to carry an angular profile. Diffusion
weighting attenuates the measured signal and limits its SNR, making independently optimized primitives prone to fitting
noise as spurious spatial structures. Both measurement noise and
slice-dependent bias can also be absorbed into the SH coefficients as false
angular variation, particularly at higher orders where the true signal is
weaker. Consequently, directly transferring independently parameterized
Gaussians from structural MRI to fetal dMRI does not adequately constrain
reconstruction in either the spatial or angular domain.

To retain the rendering advantages of Gaussian primitives without fitting
their angular attributes independently, we introduce \textsc{Angel},
\textbf{A}nchor-structured \textbf{N}eural \textbf{G}aussians for
\textbf{E}xplicit fetal diffusion MRI reconstruction with cross-subject
\textbf{L}earning. \textsc{Angel} builds on Scaffold-GS~\cite{scaffold}, whose anchors reduce redundant Gaussians while supporting view-adaptive rendering of natural scenes. In fetal dMRI, however, primitive redundancy is not the central
difficulty. The challenge is to estimate a spatial-angular signal from
noisy, motion- and intensity-corrupted slices without allowing individual Gaussians to
explain unreliable measurements in isolation. We therefore rethink
anchors as a means of structuring what is shared within each
reconstruction and what can be learned across subjects.

Within each subject, a sparse set of anchor points with learnable locations, parameters and latent features encodes an individual subject anatomy.
%carries learnable latent features. 
Each anchor supports multiple compact
signal Gaussians and one broad, slice-conditioned bias Gaussian. 
%Their common spatial support associates the bias field with the reconstructed anatomy, while 
Their different numbers and scales reflect the distinct
spatial frequencies of the two fields. More importantly, a neural
decoder shared across anchors and subjects maps subject- and anchor-specific latent features to the signal Gaussians and their SH
coefficients. Angular profiles are thus constrained by a decoder learned from measurements throughout the spatial domain and subjects, rather than fitted
independently at each primitive, reducing their tendency to absorb noise
and slice-dependent bias as spurious angular structure. This separation
between subject-specific anchors and a shared decoder also enables joint
optimization across subjects. 
%Each subject retains its own anchors and acquisition parameters. As similar local diffusion profiles occur across subjects, measurements from the entire cohort can constrain the same decoders, strengthening higher-order angular estimation without clean reference volumes or inter-subject correspondence. 
Our contributions can be summarized as follows:
\begin{itemize}
  \item We introduce \textsc{Angel}, to our knowledge the first explicit
  Gaussian framework for fetal dMRI SVR. It decodes SH coefficients from anchor features through
  a shared decoder, constraining angular profiles jointly rather than fitting independently to noisy, motion-corrupted measurements.
  \item We propose a cross-subject learning strategy that jointly optimizes
  subject-specific scaffolds and cohort-shared decoders. Sharing recurring
  angular patterns across subjects strengthens higher-order SH reconstruction. Additionally, we propose a 
  cohort-normalized latent reparameterization to stabilise transfer to unseen
  subjects.
  \item We propose an anchor-coupled bias representation that assigns one
  broad, slice-conditioned bias Gaussian to each anchor, decoupling smooth low-frequency
  intensity modulation from the reconstructed anatomy.

\item Experiments on simulated neonatal and in vivo fetal data
demonstrate more accurate spatial and angular reconstruction than
classical, implicit neural, and independently parameterized Gaussian
baselines.

\end{itemize}

\section{Related Work}
\label{sec:related_work}
Classical fetal SVR alternates between
slice registration and grid-based super-resolution, progressively refining
slice poses and the reconstructed
volume~\cite{svrtk,classic-1,classic-2}. For fetal dMRI, SH-based methods
combine a spatial voxel grid with spherical harmonic coefficients to
reconstruct measurements scattered across spatial locations and diffusion
directions~\cite{diffusion-mri-recon, diffusion-mri-recon-shard}.

With the development of deep learning~\cite{intro_deep1,intro_deep2,intro_deep3,intro_deep4}, neural representations replace the voxel grid with a continuous signal
model~\cite{inr1,inr2}. NeSVoR optimizes a coordinate-based INR jointly with slice motion, bias fields, and uncertainty through a
differentiable acquisition model~\cite{nesvor}. SHINR extends this approach
to fetal dMRI by predicting spatially varying SH coefficients, enabling
continuous spatial-angular reconstruction~\cite{shinr}.
Gaussian-based methods instead represent fetal structural MRI using
explicit, spatially localized
primitives~\cite{svr-gaussian1,svr-gaussian2}. \textit{Fast and Explicit} exploits
Gaussian convolution to incorporate the acquisition PSF analytically,
avoiding Monte Carlo integration during
optimization~\cite{svr-gaussian-fast}. 

Learning-based SVR methods learn reusable predictors from training
cohorts. SVoRT predicts slice poses through iterative Transformer-based
refinement~\cite{svort}, while SVRMamba combines sequence-aware motion
estimation with learned volumetric interpolation~\cite{svr4-mamba}.
SAFFRON removes the need for ground-truth volumes through self-supervised
training, but still learns reconstruction networks before applying them
to new subjects~\cite{svr3-label-free}. These methods therefore acquire
cross-subject information in a separate training stage. Dannecker et al.\
instead use self-supervised meta-learning to obtain INR initializations
for subsequent subject-specific optimization~\cite{meta-inr}.
\textsc{Angel} incorporates cross-subject sharing directly into reconstruction.
The same formulation supports subject-specific optimization for a single
acquisition and joint reconstruction through shared decoders when two
or more subjects are available, without requiring a separate pretraining
cohort. This flexibility extends cohort learning to spatial-angular
fetal dMRI while retaining subject-specific anatomy.

\section{Preliminary and Problem Formulation}
\label{sec:problem}

\subsection{Spatial-Angular Diffusion Signal}
\label{sec:spatial_angular_signal}

For a fixed diffusion weighting, the motion-free diffusion signal of subject
$j$ is represented as a continuous field
%\begin{equation}
$$V_j:\mathbb{R}^{3}\times\mathbb{S}^{2}\rightarrow\mathbb{R},$$
%\end{equation}
where $\mathbf{x}\in\mathbb{R}^{3}$ denotes an anatomical
location and $\mathbf{g}\in\mathbb{S}^{2}$ denotes a unit
diffusion-encoding direction. Unlike structural MRI, which associates one
scalar intensity with each spatial location, dMRI defines a spherical signal
%profile 
$V_j(\mathbf{x},\cdot)$ that characterizes the directional dependence
of water diffusion within the underlying tissue. The angular profile of the dMRI signal is represented using a real SH
basis:
\begin{equation}
V_j(\mathbf{x},\mathbf{g})
\approx
\mathbf{c}_j(\mathbf{x})^{\mathsf T}\mathbf{Y}(\mathbf{g}),
\label{eq:sh_signal}
\end{equation}
where $\mathbf{Y}(\mathbf{g})\in\mathbb{R}^{M}$ collects the SH basis
functions up to degree $L$, and
$\mathbf{c}_j(\mathbf{x})\in\mathbb{R}^{M}$ contains their spatially varying
coefficients, with $M=(L+1)(L+2)/2$~\cite{dmri-shfit}. Since the diffusion signal is assumed
to be antipodally symmetric,
$V_j(\mathbf{x},\mathbf{g})=V_j(\mathbf{x},-\mathbf{g})$, only even SH
degrees are retained. Reconstructing the diffusion signal is therefore
equivalent to estimating the continuous SH coefficient field
$\mathbf{c}_j(\mathbf{x})$.

\subsection{Motion-Corrupted Slice Acquisition}
\label{sec:slice_acquisition}

The measurements of subject $j$ consist of motion-corrupted slices
$\{y_{j,n,p}\}$, where $n$ indexes the slice and $p$ indexes a voxel within
that slice. Let $\mathbf{x}^{\mathrm{sc}}_{j,n,p}$ denote the voxel location
in scanner coordinates. Since each slice is acquired rapidly, intra-slice motion is assumed to be negligible, and the fetal pose during the acquisition of a single slice \(n\) is represented by a rigid transformation $T_{j,n}$ from anatomical to scanner coordinate, with translation $\mathbf{t}_{j,n}$ and rotation $R_{j,n}$:
\begin{equation}
T_{j,n}(\mathbf{x})
=
R_{j,n}\mathbf{x}+\mathbf{t}_{j,n},
\;\;
R_{j,n}\in\mathrm{SO}(3),\; \mathbf{t}_{j,n}\in\mathbb{R}^{3}.
\label{eq:slice_transform}
\end{equation}
The corresponding anatomical coordinate and motion-corrected diffusion
direction are:
\begin{equation}
\mathbf{x}_{j,n,p}
=
T_{j,n}^{-1}
\left(\mathbf{x}^{\mathrm{sc}}_{j,n,p}\right),
\qquad
\mathbf{g}'_{j,n}
=
R_{j,n}^{-1}\mathbf{g}_{j,n}.
\label{eq:motion_corrected_coordinates}
\end{equation}
Fetal motion therefore changes both the spatial location at which the signal
is sampled and the diffusion direction along which it is evaluated.

% A measured voxel integrates the underlying signal over the finite slice
% point spread function (PSF) and is additionally affected by spatially
% varying intensity bias and global slice scaling. 
Defining the motion-corrected diffusion signal as $V_{j,n}(\mathbf{x})
=
V_j(\mathbf{x},\mathbf{g}'_{j,n}),
$ the simulated measurement through the acquisition process is given by:
\begin{equation}
\widehat{y}_{j,n,p}
=
\gamma_{j,n}
\left[
P_{j,n}*
\left(
B_{j,n}V_{j,n}
\right)
\right]
(\mathbf{x}_{j,n,p}),
\label{eq:slice_acquisition}
\end{equation}
where $P_{j,n}$ denotes the slice-dependent point spread function (PSF) rotated by the estimated slice motion. This PSF models spatial
blurring related to resolution of acquired MRI slices as a 3D Gaussian with covariance $\boldsymbol{\Sigma}^{P}_{j,n}=R_{j,n}\boldsymbol{\Sigma}_jR_{j,n}^T$ where $\boldsymbol{\Sigma}_j$ is diagonal covariance of the PSF in the scanner space.
$B_{j,n}(\mathbf{x})$ is a slice-dependent multiplicative bias field
modeling smooth spatial intensity inhomogeneity; $\gamma_{j,n}$ is a global
slice-wise scaling factor accounting for signal attenuation; and $*$ denotes convolution.

Minimizing the discrepancy between the acquired measurements \(y_{j,n,p}\) and their simulated counterparts \(\widehat{y}_{j,n,p}\) enables reconstruction of the underlying diffusion signal \(V_j(\mathbf{x},\mathbf{g})\) from motion- and acquisition-corrupted slices.

% Given a cohort of acquired slices
% $\{\mathcal{D}_j\}_{j=1}^{N_{\mathrm{sub}}}$, the reconstruction problem is
% to recover each subject-specific field
% $V_j(\mathbf{x},\mathbf{g})$ together with its slice motion and intensity
% nuisance parameters. The following section addresses this inverse problem by
% representing each $V_j$ with a subject-specific Gaussian scaffold while
% sharing its neural decoder across the cohort.
\section{Method}

\begin{figure*}
\includegraphics[width=\textwidth]{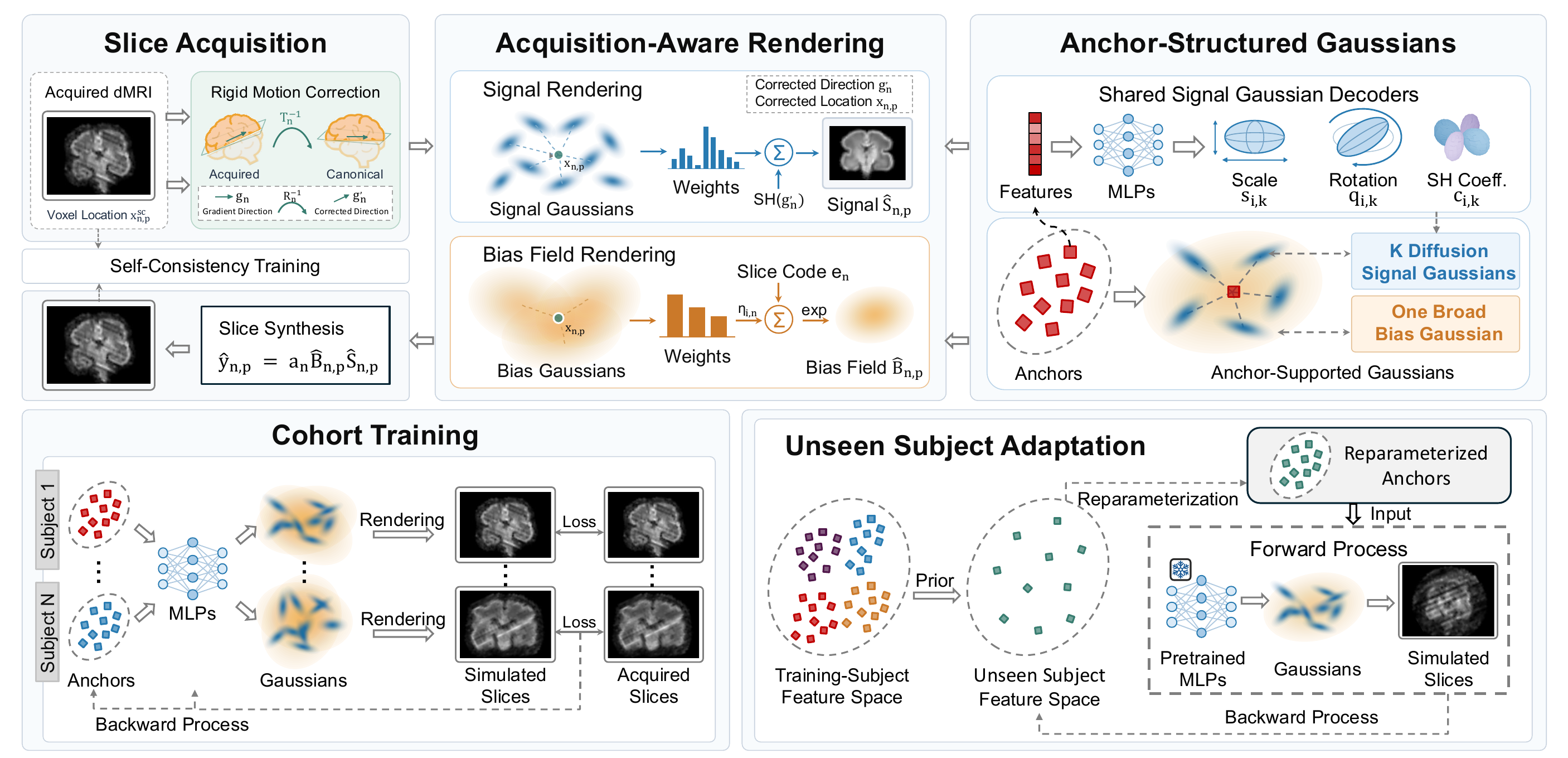}
\caption{Overview of \textsc{Angel}.
\textbf{Top:} (Left) Motion correction provides anatomical
locations and reoriented diffusion directions.
(Right) Shared MLPs decode anchor features into signal
Gaussians, while each anchor carries a slice-conditioned
bias Gaussian.
(Middle) These representations render diffusion signals,
which are passed through the acquisition forward model
to synthesize motion-corrupted slices for
self-consistency optimization.
\textbf{Bottom:} Decoders are shared across subjects during
cohort training and frozen during adaptation to unseen
subjects, while anchors remain subject-specific.}
\label{main_fig}
\end{figure*}

% \subsection{Overview: Subject-Factorized Neural Gaussian Field}
\subsection{Overview}
\label{sec:overview}

Given a cohort of $N_{\mathrm{sub}}$ motion-corrupted fetal dMRI acquisitions
$\{\mathcal{D}_j\}_{j=1}^{N_{\mathrm{sub}}}$, the proposed framework learns a mapping $\mathcal{G}$ that jointly
predicts the subject-specific spatial--angular fields
$\{V_j(\mathbf{x},\mathbf{g})\}$ introduced in
Sec.~\ref{sec:spatial_angular_signal}. Each reconstruction is factorized into
a subject-specific Gaussian scaffold $\boldsymbol{\Phi}_j$ and a neural
decoder $\boldsymbol{\Theta}$ shared across the cohort:
\begin{equation}
V_j(\mathbf{x},\mathbf{g})
=
\mathcal{G}
\left(
\mathbf{x},
\mathbf{g};
\boldsymbol{\Phi}_j,
\boldsymbol{\Theta}
\right),
\label{eq:subject_factorization}
\end{equation}
where $\boldsymbol{\Phi}_j$ contains the anchor locations, latent feature vectors and additional learnable parameters (Sec.~\ref{sec:signal_scaffold}). The decoder $\boldsymbol{\Theta}$ maps each anchor feature vector to multiple Gaussian primitives which parametrise SH coefficients $\mathbf{c}_j(\mathbf{x})$ of the dMRI signal, thereby providing a continuous representation of
$V_j(\mathbf{x},\mathbf{g})$.

The decoder $\boldsymbol{\Theta}$ is shared across anchors and subjects, thus regularising estimation of SH coefficients spatially within the same subject, while 
%This factorization shares information at two complementary levels. 
%Within each subject, the common decoder couples the Gaussian primitives and regularizes the estimation of the SH coefficient fields.
%$\mathbf{c}_j(\mathbf{x})$ under low signal-to-noise ratio. 
%Across subjects, the same decoder $\boldsymbol{\Theta}$ is jointly constrained by all
%acquisitions, 
allowing recurrent spatial--angular characteristics to inform
the reconstruction of the entire cohort. Subject-specific anatomy remains
encoded by $\boldsymbol{\Phi}_j$, and the anchors are neither shared nor
spatially aligned across subjects.

In addition to the signal field, the scaffold supports the slice-dependent
Gaussian representation of bias $B_{j,n}$. The slice motion $T_{j,n}$ and intensity
scaling $\gamma_{j,n}$ are represented by independent slice-specific parameters. Together with the signal field $V_j(\mathbf{x},\mathbf{g})$, these components
are incorporated into the forward model in
Eq.~\eqref{eq:slice_acquisition} and jointly estimated from the acquired
slices.

\subsection{Anchor-Structured Spatial-Angular Gaussians}
\label{sec:signal_scaffold}
%Independently optimizing Gaussian geometry and SH coefficients overfits the low-SNR measurements of fetal dMRI by absorbing noise into spurious spatial and angular variations. 
%We 
%therefore 
%generate attributes of the Gaussian primitives from compact anchor features through the shared neural decoder $\boldsymbol{\Theta}$.
%, constraining both spatial and angular variations to a common data-adaptive parameterization. 
For each subject $j$, the diffusion signal $V_j$ is organized by a set of anchors $\boldsymbol{\Phi}_j = \{\mathcal{A}_{j,i}\}_{i=1}^{N_j}$ distributed in its anatomical
space, in order to provide a compact Gaussian representation designed to prevent overfitting to noise and artefacts.
%$\boldsymbol{\Phi}_j = \{\mathcal{A}_{j,i}\}_{i=1}^{N_j}$  
Each anchor supports $K$ signal Gaussians and one slice-dependent bias Gaussian, and is defined by learnable parameters $\mathcal{A}_{j,i} = \{\mathbf{a}_{j,i},\mathcal{S}_{j,i},\mathcal{B}_{j,i}\}$, where $\mathbf{a}_{j,i}$ is the anchor location, $\mathcal{S}_{j,i}$ are the parameters encoding the dMRI signal, and $\mathcal{B}_{j,i}$ encode slice-wise bias fields.
The signal Gaussians are defined by parameters
\begin{equation}
\mathcal{S}_{j,i}
=
\left\{
\rho_{j,i},
\mathbf{f}_{j,i},
\{\mathbf{o}_{j,i,k}\}_{k=1}^{K}
\right\},
\end{equation}
where scalar $\rho_{j,i}$ controls anchor's spatial extent,
$\mathbf{f}_{j,i}$ is a compact latent feature vector, and
$\mathbf{o}_{j,i,k}$ denotes the offset of $k$th signal Gaussian primitive. The
centers of the signal Gaussian primitives are given by:
\begin{equation}
\boldsymbol{\mu}^{S}_{j,i,k}
=
\mathbf{a}_{j,i}
+
\rho_{j,i}\cdot\mathbf{o}_{j,i,k}.
\end{equation}
%where superscript $S$ denotes  variables defining the dMRI signal.
Their scales and  rotations, and SH coefficients parametrising their direction-dependent responses  are
decoded from the compact anchor feature vector:
\begin{equation}
\left\{
\mathbf{s}^{S}_{j,i,k},
\mathbf{q}^{S}_{j,i,k},
\mathbf{c}_{j,i,k}
\right\}_{k=1}^{K}
=
\boldsymbol{\Theta}(\mathbf{f}_{j,i}).
\end{equation}
The shared decoder $\boldsymbol{\Theta}$ is implemented by two multi-layer perceptrons, one predicting scales $\mathbf{s}^{S}_{j,i,k}$ and rotations $\mathbf{q}^{S}_{j,i,k}$, which together define the covariance matrices $\boldsymbol{\Sigma}^{S}_{j,i,k}$ of the Gaussian primitives \cite{svr-gaussian-fast}, and the other SH coefficients $\mathbf{c}_{j,i,k}$. 
%where $D_{\boldsymbol{\Theta}}$ collectively denotes the shared scale,rotation, and SH decoders. 
%Each coefficient vector
%$\mathbf{c}_{j,i,k}$ parameterizes an even-order SH signal and therefore
%defines the direction-dependent response of the primitive.
%The same anchor scaffold is also used to represent the bias field
%Since intensity inhomogeneity arises from the interaction between the underlying anatomy and scanner sensitivity, we couple its spatial support to the reconstructed signal rather than parameterizing it with an independent set of primitives. Specifically, while each anchor supports multiple compact signal Gaussians, it additionally supports a 
The single bias Gaussian is defined by parameters
\begin{equation}
\mathcal{B}_{j,i}
=
\left\{
\boldsymbol{\mu}^{B}_{j,i},
\mathbf{s}^{B}_{j,i},
\mathbf{q}^{B}_{j,i},
\mathbf{z}^{B}_{j,i}
\right\},
\end{equation}
where $\boldsymbol{\mu}^{B}_{j,i}$, $\mathbf{s}^{B}_{j,i}$, and $\mathbf{q}^{B}_{j,i}$ denote its center, scale, and rotation, respectively, and $\mathbf{z}^{B}_{j,i}\in\mathbb{R}^{8}$ is a compact embedding of slice-dependent bias variation at the same spatial location. The bias Gaussian shares its centre with the anchor location $\boldsymbol{\mu}^{B}_{j,i} = \mathbf{a}_{j,i}$, and has  a substantially larger spatial scale, $|\mathbf{s}^{B}_{j,i}|^{2} \gg |\mathbf{s}^{S}_{j,i,k}|^{2}$, yielding a smoother, lower-frequency representation of intensity variation.

\subsection{Anchor-Based Acquisition-Aware Rendering}
For each acquired voxel, the same local support is determined for both diffusion signal and bias estimation in the shared anchor space. The motion-corrected coordinate $\mathbf{x}_{j,n,p}$ retrieves the $K_{\mathrm{nn}}$ nearest anchors, from which all associated signal and bias primitives are used for rendering. 
%Since each anchor supports $K$ signal Gaussians and one bias Gaussian, the same neighborhood naturally provides the spatial support for both diffusion signal and bias estimation.

\noindent\textbf{Diffusion signal rendering.}
Each neighboring signal primitive is first evaluated along the motion-corrected diffusion direction $\mathbf{g}'_{j,n}$, with its directional response determined by the SH coefficients:
\begin{equation}
v_{j,i,k,n}
=
\mathbf{c}_{j,i,k}^{\mathsf T}
\mathbf{Y}(\mathbf{g}'_{j,n}).
\end{equation}

Following the analytic PSF formulation of \textit{Fast-and-Explicit} \cite{svr-gaussian-fast}, the finite acquisition resolution is incorporated by convolving each signal Gaussian with the slice PSF, by adding their covariances:
\begin{equation}
\boldsymbol{\Sigma}_{j,i,k,n}
=
\boldsymbol{\Sigma}^{S}_{j,i,k}
+
\boldsymbol{\Sigma}^{P}_{j,n}.
\end{equation}
Let $G_{j,i,k,n}(\mathbf{x})$ denote the PSF-blurred signal Gaussian with covariance $\boldsymbol{\Sigma}_{j,i,k,n}$. The diffusion signal at the acquired voxel is then rendered by normalized local aggregation:
\begin{equation}
\label{eq:gaussian-interplation}
\widehat{V}_{j,n,p}
=
\frac{
\displaystyle
\sum_{i\in\mathcal{N}_{j}(\mathbf{x}_{j,n,p})}
\sum_{k=1}^{K}
G_{j,i,k,n}(\mathbf{x}_{j,n,p})
v_{j,i,k,n}
}{
\displaystyle
\sum_{i\in\mathcal{N}_{j}(\mathbf{x}_{j,n,p})}
\sum_{k=1}^{K}
G_{j,i,k,n}(\mathbf{x}_{j,n,p})
+
\varepsilon
},
\end{equation}
where $\mathcal{N}_{j}(\mathbf{x}_{j,n,p})$ denotes the set of neighborhood $K_{\mathrm{nn}}$ and $\varepsilon$ is a small constant for numerical stability.

\noindent\textbf{Slice-conditioned bias rendering.} The same anchor neighborhood is used to render the bias field. To capture slice-dependent intensity variation, each slice $n$ is assigned a learnable embedding $\mathbf{e}_{j,n}\in\mathbb{R}^{8}$. Its interaction with the local bias embedding determines the slice-specific bias response:
\begin{equation}
\eta_{j,i,n}
=
\left(\mathbf{z}^{B}_{j,i}\right)^{\mathsf T}
\mathbf{e}_{j,n}.
\end{equation}
The neighboring bias Gaussian responses $\eta_{j,i,n}$ are aggregated using the same Gaussian-weighted interpolation as in Eq.~\eqref{eq:gaussian-interplation}. This produces smooth spatially varying log-bias field, which is exponentiated to obtain the positive multiplicative bias $\widehat{B}_{j,n,p}$. 
%The spatially distributed bias embeddings capture variation across the reconstructed volume, while the slice embeddings account for acquisition-dependent intensity changes.

\noindent\textbf{Slice rendering.} Finally, the synthesized voxel signal is given by
\begin{equation}
\widehat{y}_{j,n,p}
=
\gamma_{j,n}\,
\widehat{B}_{j,n,p}\,
\widehat{V}_{j,n,p}.
\end{equation}
Eqn.~\ref{eq:slice_acquisition} is approximated by rendering signal and bias separately to avoid the quadratic number of Gaussian interactions that would arise from explicitly convolving their product.

\subsection{Cross-Subject Training and Adaptation}
\label{sec:cross_subject}
Conventional subject-specific reconstruction independently learns a representation from each acquisition, requiring common spatial-angular structure to be repeatedly estimated from noisy measurements. Our formulation instead separates the subject-specific signal scaffold $\boldsymbol{\Phi}_j$  from the shared neural decoder $\boldsymbol{\Theta}$.
%, allowing the decoder to be shared across subjects while each scaffold remains specific to its anatomy. 
This supports two complementary settings: joint cohort training and adaptation of the learned decoder to an unseen subject.

\noindent\textbf{Cohort training.}
Let 
%$\mathcal{D}_j$ denote the acquired measurements of subject $j$, and define the pooled cohort dataset as
$
\mathcal{D}
=
\textstyle\bigcup_{j=1}^{N_{\mathrm{sub}}}\mathcal{D}_j$ define the pooled cohort dataset.
%, where $N_\mathrm{sub}$ is the number of subjects. 
At each iteration, a mini-batch $\mathcal{M}$ is sampled from $\mathcal{D}$ and may contain measurements from multiple subjects. For each subject $j$ represented in the mini-batch, we denote
$\mathcal{M}_j
=
\mathcal{M}\cap\mathcal{D}_j$ as its corresponding subset. The cohort objective is defined as:
\begin{equation}
\mathcal{L}_{\mathrm{C}}
=
\mathbb{E}_{\mathcal{M}\sim\mathcal{D}}
\!\left[
\sum_{j\in\mathcal{J}(\mathcal{M})}
\frac{|\mathcal{M}_j|}{|\mathcal{M}|}
\mathcal{L}_j(\boldsymbol{\Theta},\boldsymbol{\Phi}_j;\mathcal{M}_j)
\right],
\label{eq:cohort_objective}
\end{equation}
where $\mathcal{J}(\mathcal{M})$ denotes the subjects represented in $\mathcal{M}$. Measurements from all subjects jointly update the shared decoder $\boldsymbol{\Theta}$, whereas each signal scaffold $\boldsymbol{\Phi}_j$ receives gradients only from its own measurements. The decoder therefore captures spatial-angular regularities shared across the cohort, while the anchors retain subject-specific anatomy. No ground-truth reconstruction or inter-subject anchor correspondence is required, and the formulation reduces to conventional subject-wise optimization when $N_{\mathrm{sub}}=1$.

\noindent\textbf{Adaptation to an unseen subject.}
For an unseen subject $u$, the cohort-pretrained decoder $\boldsymbol{\Theta}^{\star}$ is frozen while a new scaffold $\boldsymbol{\Phi}_u$ is optimized from its acquired measurements. The anchor feature vectors $\mathbf{f}_{u,i}$ therefore adapt within the cohort-learned latent space. Other subject-specific acquisition parameters are optimized as in subject-wise reconstruction.

We observe that the dispersion of the learned anchor features increases with the number of subjects used for cohort training, as reflected by an increasing standard deviation over the training anchors. Consequently, the same learning rate in the raw feature space corresponds to different relative update scales for models pretrained on different cohort sizes. 

To account for this variation, we introduce Cohort-Normalized Latent Reparameterization (CNLR), which parameterizes unseen-subject anchor features relative to the latent statistics of the pretrained cohort. Specifically, we compute the element-wise mean $\boldsymbol{\mu}_f$ and standard deviation $\boldsymbol{\sigma}_f$ of the training anchor features and parameterize:
\begin{equation}
\label{eq:repa}
\mathbf{f}_{u,i}
=
\boldsymbol{\mu}_f
+
\boldsymbol{\sigma}_f
\odot
\boldsymbol{\xi}_{u,i},
\end{equation}
where $\boldsymbol{\xi}_{u,i}$ is the learnable normalized latent coordinate. Adaptation is thus performed in a normalized latent space whose scale is independent of the dispersion of the pretrained feature space, allowing the same optimization schedule to be used across models trained with different numbers of subjects.

Once $\boldsymbol{\Theta}^{\star}$, $\boldsymbol{\mu}_f$, and $\boldsymbol{\sigma}_f$ are stored, adaptation requires only the measurements of the unseen subject and no access to the original training cohort.

\subsection{Reconstruction Objective}
\label{sec:optimization}

For each subject $j$, the reconstruction objective is evaluated over the
corresponding mini-batch subset $\mathcal{M}_j$ defined in
Sec.~\ref{sec:cross_subject}. Following previous work \cite{nesvor}, each slice $n$ is
assigned a learnable variance $\nu_{j,n}^{2}$ shared by all of its voxels.
The data-fidelity term is formulated as a Gaussian negative log-likelihood:
\begin{equation}
\mathcal{L}_{\mathrm{NLL}}^{j}
=
\frac{1}{|\mathcal{M}_j|}
\sum_{(n,p)\in\mathcal{M}_j}
\left[
\frac{r_{j,n,p}^{2}}{2\nu_{j,n}^{2}}
+
\frac{1}{2}\log\nu_{j,n}^{2}
\right],
\label{eq:nll_loss}
\end{equation}
where $r_{j,n,p}=y_{j,n,p}-\widehat{y}_{j,n,p}$ denote the reconstruction residuals. The learned variance reduces the influence of severely corrupted slices,
while the logarithmic term prevents unbounded uncertainty inflation.

Additional regularization is imposed on the bias field, Gaussian scales, and
slice motion. Following \cite{nesvor}, the bias regularizer $\mathcal{R}_{B}^{j}$ penalizes the
rendered log-bias toward zero to prevent the bias representation from
absorbing diffusion signal variation. Similar to \cite{svr-gaussian-fast}, the Gaussian primitive scale regularizers
$\mathcal{R}_{s}^{B,j}$ and $\mathcal{R}_{s}^{V,j}$ impose minimum spatial
scales on the bias and signal primitives, respectively, with a larger lower
bound for the bias Gaussians to preserve their smooth, low-frequency
behavior. The motion regularizer $\mathcal{R}_{T}^{j}$ discourages
implausibly large slice transformations \cite{nesvor}. The complete subject-specific objective is
\begin{equation}
\begin{aligned}
\mathcal{L}_{j}
={}&
\mathcal{L}_{\mathrm{NLL}}^{j}
+
\lambda_{B}\mathcal{R}_{B}^{j}
+
\lambda_{sB}\mathcal{R}_{s}^{B,j} \\
&+
\lambda_{sV}\mathcal{R}_{s}^{V,j}
+
\lambda_{T}\mathcal{R}_{T}^{j}.
\end{aligned}
\label{eq:total_objective}
\end{equation}
During cohort training, $\mathcal{L}_{j}$ is substituted into
Eq.~\eqref{eq:cohort_objective}, jointly optimizing the shared decoder, the subject-specific signal and bias scaffolds, 
slice transformations, intensity scaling factors, and uncertainty estimates.
For unseen-subject adaptation, the same reconstruction objective is used
while keeping the pretrained decoder fixed.

% \subsection{Reconstruction Objective}
% \label{sec:objective}

% For each subject, the reconstruction objective is defined by a Gaussian
% negative log-likelihood between the synthesized and acquired measurements.
% A learnable variance is assigned to each slice and shared by all of its
% voxels, reducing the influence of slices affected by severe corruption.
% Additional regularization terms constrain the rendered log-bias magnitude,
% enforce minimum spatial scales for the signal and bias primitives, and
% discourage implausibly large motion updates. The subject-specific objective
% is
% \begin{equation}
% \mathcal{L}_{j}
% =
% \mathcal{L}^{j}_{\mathrm{NLL}}
% +
% \lambda_B\mathcal{R}^{j}_{B}
% +
% \lambda_{sB}\mathcal{R}^{B,j}_{s}
% +
% \lambda_{sV}\mathcal{R}^{V,j}_{s}
% +
% \lambda_T\mathcal{R}^{j}_{T}.
% \end{equation}
% Substituting this objective into
% \eqref{eq:cohort_objective} jointly optimizes the cohort-shared decoder and
% all subject-specific scaffolds, bias representations, slice transformations,
% scaling factors, and uncertainty estimates using only the acquired
% motion-corrupted slices.
\section{Experiments}
\label{sec:experiments}

\subsection{Datasets}
%Experiments are conducted on one in-vivo fetal dataset and one simulated neonatal dataset.

\noindent\textbf{In-vivo fetal dataset.}
Diffusion MRI scans of nine healthy fetal subjects were acquired on a 3T Philips Achieva scanner. Three scans were acquired using a clinical protocol with 32 diffusion directions at $b=750~\mathrm{s/mm^2}$, with resolution $2\times2\times3.5~\mathrm{mm^3}$. The remaining six were acquired using the developing Human Connectome Project (dHCP) ~\cite{diffusion-mri-recon-shard,dhcp2,dhcp1-diffusion,dhcp2-diffusion} 
%~\cite{dhcp,dhcp2,dhcp3,dhcp1-diffusion,dhcp2-diffusion} 
with isotropic resolution $2\times2\times2~\mathrm{mm^3}$ . In our experiments we used the highest $b=1000~\mathrm{s/mm^2}$ shell with 80 diffusion gradient directions. As no motion-free reference is available, five visually uncorrupted diffusion-weighted volumes are held out for self-consistency evaluation.

\noindent\textbf{Simulated neonatal dMRI dataset.}
Twenty high-quality dHCP neonatal reconstructions~\cite{diffusion-mri-recon-shard} spanning 26-44 weeks post-menstrual age are used as pseudo-ground truth. The $b=1000~\mathrm{s/mm^2}$ shell is selected, comprising 88 diffusion directions at $1.5~\mathrm{mm}$ isotropic resolution. Fetal-like slice acquisitions are simulated from these reference volumes
% \noindent\textbf{Simulated neonatal dMRI template dataset.}
% A neonatal dMRI template constructed from dHCP subjects between 33 and 44 weeks post-menstrual age was additionally used as a reference. The $b=1000~\mathrm{s/mm^2}$ shell was selected, comprising 80 diffusion directions at $1.0~\mathrm{mm}$ isotropic resolution. The same fetal-like acquisition simulation was applied to the template.
%\subsection{Simulation Protocol}
%Fetal-like dMRI slice acquisitions are simulated 
along five dimensions: spatial resolution, slice motion, bias field, noise level, and signal artifacts. 
The reference volumes are first resampled to $2\times2\times2~\mathrm{mm^3}$ isotropic resolution. Inter-slice motion is simulated by independently sampling translations and rotations from $\mathcal{U}(-3,3)~\mathrm{mm}$ and $\mathcal{U}(-6,6)^\circ$, respectively. To model slice-wise intensity inhomogeneity, an independent bias field is generated for each slice using MONAI \cite{monai} and applied to the corresponding signal. Assuming a shared noise level across diffusion shells, the $b=0$ SNR is sampled from $\mathcal{U}(20,40)$, and the corresponding Rician noise is applied to the selected $b=1000~\mathrm{s/mm^2}$ shell. Finally, signal dropout is simulated by introducing void artifacts in $5\%$ of the acquired measurements.

\subsection{Evaluation and Implementation Details}

\noindent\textbf{Evaluation metrics.} For the simulated neonatal dataset with motion-free ground truth, the spatial domain accuracy is measured using PSNR and NCC of the reconstructed DWI volumes against the corresponding reference images. To evaluate accuracy in angular domain, SH are fitted to the reconstructed and reference signals, and the PNSR and NCC metrics are reported for SH coefficients separately for each anisotropic degree \(l=2,4\). Angular accuracy is further assessed by comparing reconstructed and reference fibre orientation distributions (FOD), which are obtained from the DWI signals by constrained spherical deconvolution \cite{csd1,csd2,csd3}. Angular metrics include full angular cross correlation (ACC) over all SH degrees, shape ACC over the anisotropic \(l=2,4\) components, and the mean peak angular error (PAE) between matched FOD peaks. For the in-vivo fetal data, which lack motion-free ground truth, evaluation is restricted to spatial DWI metrics between acquired and predicted signals on the held-out DWI stacks of slices.

%Reconstruction is evaluated at the DWI and angular levels. DWI accuracy is measured using PSNR and NCC against the corresponding reference images. For the simulated neonatal datasets with motion-free ground truth, SH coefficients are fitted to the reconstructed and reference signals, and the same metrics are reported separately for degrees \(l=0,2,4\). Angular accuracy is further assessed using fibre orientation distribution (FOD) \cite{csd1,csd2,csd3} angular correlation coefficient (ACC), including full ACC over all SH degrees and shape ACC over the anisotropic \(l=2,4\) components, together with the mean peak angular error (PAE) between matched FOD peaks. For the in-vivo fetal data, which lack motion-free ground truth, evaluation is restricted to DWI metrics on the held-out measurements.

\noindent\textbf{Implementation details.} For the fetal datasets, we use 4,000 anchors with five signal Gaussians per anchor and real even-order SH basis functions up to \(L=4\). The minimum signal and bias scales are \(s^S=2~\mathrm{mm}\) and \(s^B=10~\mathrm{mm}\), respectively. Anchor features are 64-dimensional, bias and slice embeddings are 8-dimensional, the anchor extent is initialised as \(\rho=2~\mathrm{mm}\), and \(30\) nearest anchors are used for rendering. For the simulated neonatal datasets, we use 10,000 anchors, \(s^S=1~\mathrm{mm}\), and \(12\) nearest anchors; all other settings are unchanged.

\subsection{Comparative Results}
\noindent\textbf{Comparison methods.} The proposed method is compared against IRTK \cite{diffusion-mri-recon} and SHINR \cite{shinr}, two representative methods for diffusion MRI slice-to-volume reconstruction. An additional Gaussian baseline is adapted from \textit{Fast-and-Explicit}~\cite{svr-gaussian-fast} by replacing the scalar Gaussian intensity with SH coefficients and incorporating the proposed bias-field module. This baseline is denoted as \textit{Plain-GS} \cite{svr-gaussian-fast}. The proposed model is trained using the cohort-training strategy described in Sec.~\ref{sec:cross_subject}. For the fetal data, the six subjects acquired at $b=1000~\mathrm{s/mm^2}$ are jointly optimized as one cohort, while the three subjects acquired at $b=750~\mathrm{s/mm^2}$ form a separate cohort. All twenty simulated neonatal subjects are jointly optimized as a single cohort.

\noindent\textbf{Reconstruction results.}
Table~\ref{tab:main_dwi_comparison} summarizes the DWI reconstruction
results. \textsc{Angel} outperformed IRTK, SHINR, and \textit{Plain-GS}
on both PSNR and NCC across the fetal and simulated neonatal datasets.
On fetal data, the respective PSNR and NCC gains were 1.21~dB and
0.0493 over IRTK, 1.01~dB and 0.0384 over SHINR, and 0.69~dB and
0.0292 over \textit{Plain-GS}. Both Gaussian-based methods also
achieved shorter reconstruction times than IRTK and SHINR.

The neonatal reference reconstructions further enabled direct evaluation
of angular fidelity. As reported in
Table~\ref{tab:main_angular_comparison}, \textsc{Angel} achieved higher
accuracy for both the second- and fourth-order SH components.
It also obtained the highest full and shape FOD ACC and the lowest
mean angular error between matched FOD peaks, demonstrating improvements
beyond DWI intensity reconstruction.

\begin{figure}[t]
    \centering
    \setlength{\tabcolsep}{1pt}
    \renewcommand{\arraystretch}{0}

    \begin{tabular}{@{}ccccc@{}}
        \scriptsize\textbf{GT} &
        \scriptsize\textbf{IRTK} &
        \scriptsize\textbf{SHINR} &
        \scriptsize\textbf{Plain-GS} &
        \scriptsize\textsc{\textbf{Angel}} \\[2pt]

        \includegraphics[width=0.19\columnwidth]{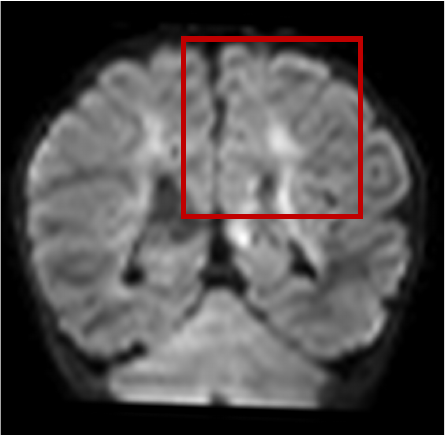} &
        \includegraphics[width=0.19\columnwidth]{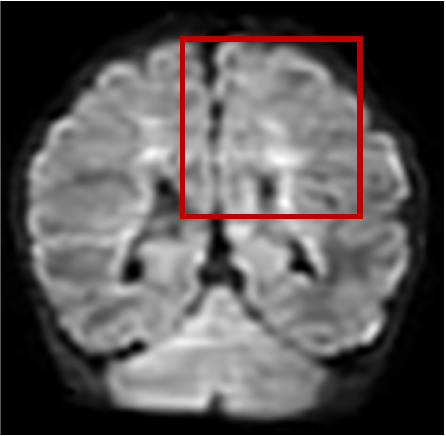} &
        \includegraphics[width=0.19\columnwidth]{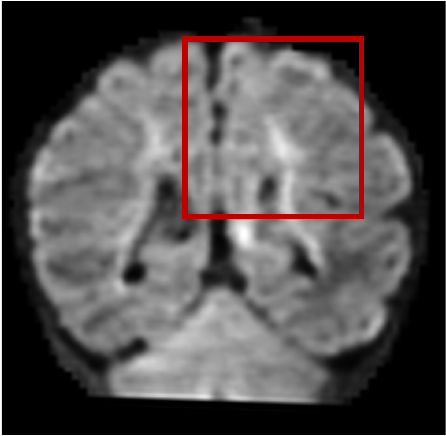} &
        \includegraphics[width=0.19\columnwidth]{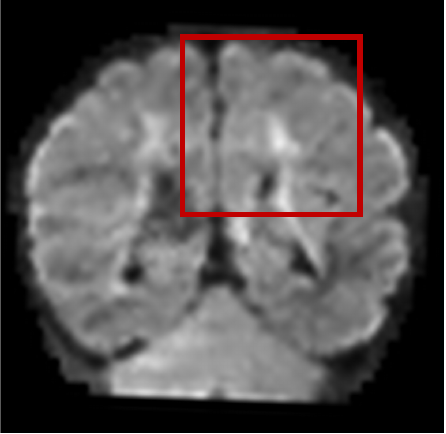} &
        \includegraphics[width=0.19\columnwidth]{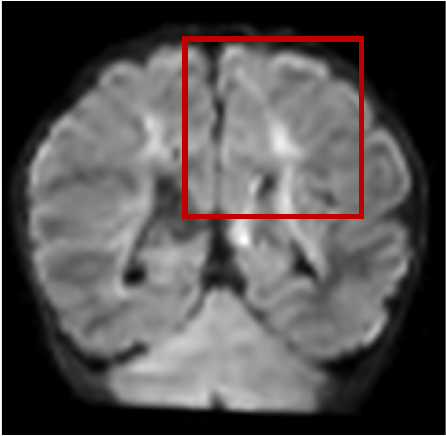} \\[2pt]

        \includegraphics[width=0.19\columnwidth]{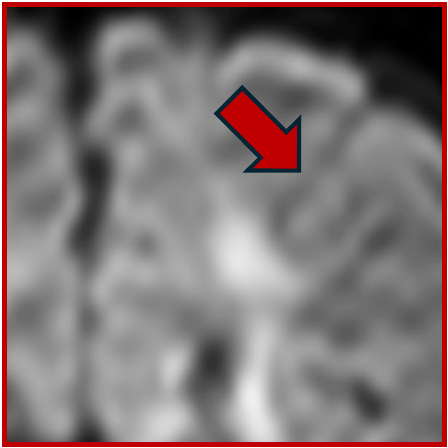} &
        \includegraphics[width=0.19\columnwidth]{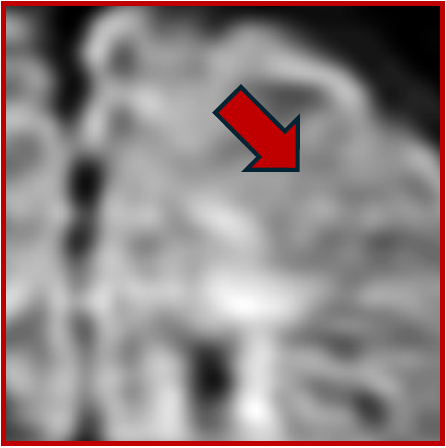} &
        \includegraphics[width=0.19\columnwidth]{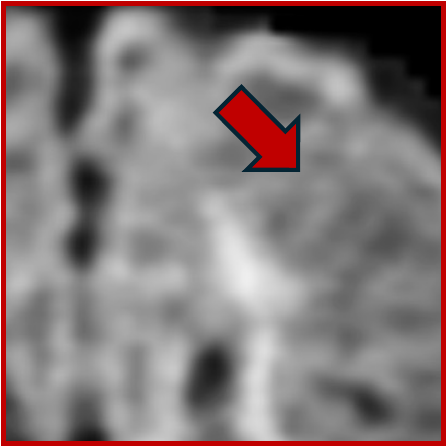} &
        \includegraphics[width=0.19\columnwidth]{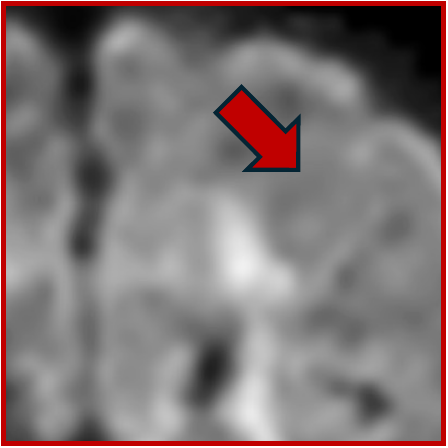} &
        \includegraphics[width=0.19\columnwidth]{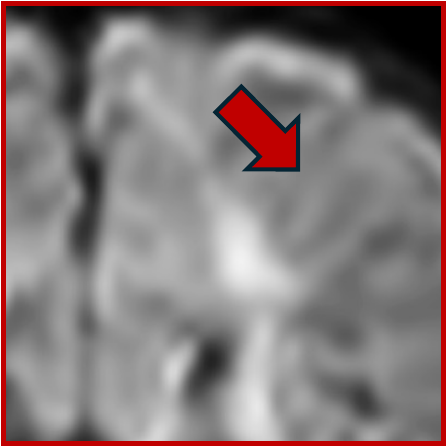}
    \end{tabular}

    \caption{Qualitative comparison of DWI reconstructions across methods on simulated neonatal data. Red arrows: complex cortical folding is more accurately reconstructed with ANGEL than with the other methods.}
    \label{fig:comparsion_neonatal_dwi}
\end{figure}

\begin{figure}[t]
    \centering
    \setlength{\tabcolsep}{1pt}
    \renewcommand{\arraystretch}{0}

    \begin{tabular}{@{}cccc@{}}
        {\fontsize{6pt}{6pt}\textsf{\textbf{IRTK}}} &
        {\fontsize{6pt}{6pt}\textsf{\textbf{SHINR}}} &
        {\fontsize{6pt}{6pt}\textsf{\textbf{Plain-GS}}} &
        \scriptsize\textsc{\textbf{Angel}} \\[2pt]

        \includegraphics[width=0.24\linewidth]{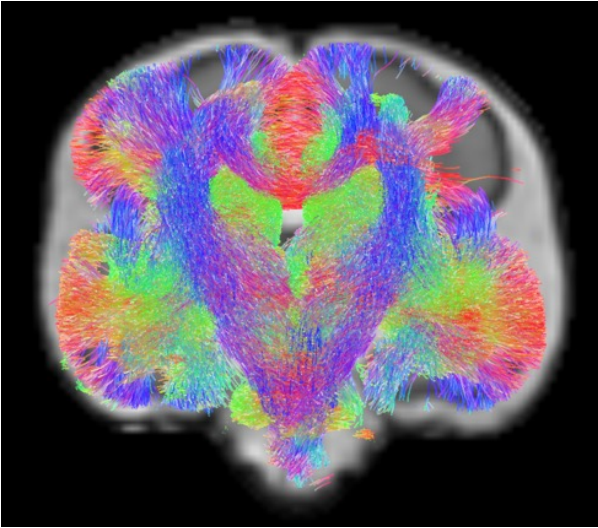} &
        \includegraphics[width=0.24\linewidth]{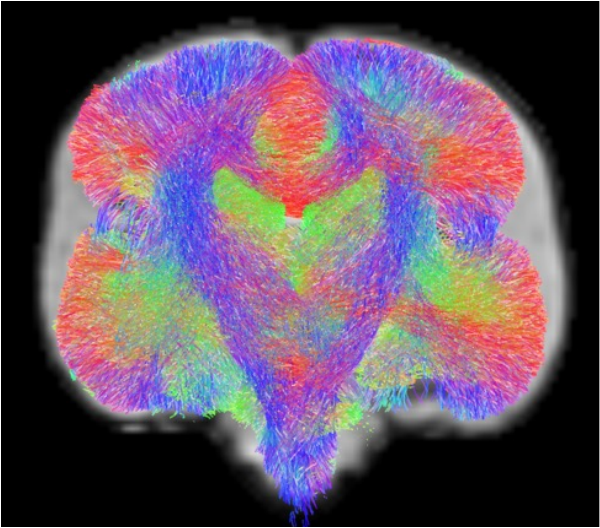} &
        \includegraphics[width=0.24\linewidth]{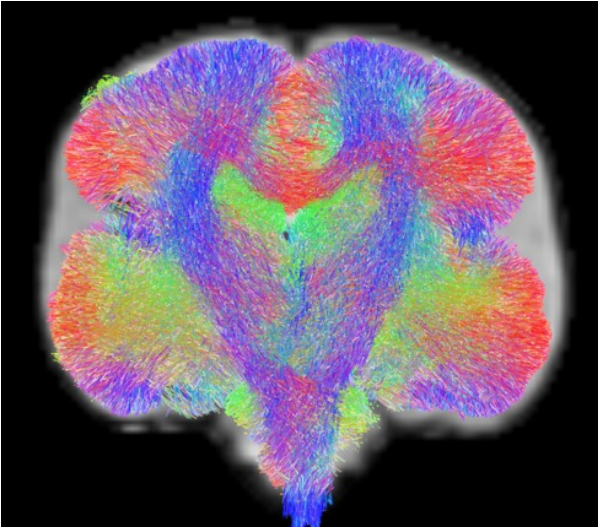} &
        \includegraphics[width=0.24\linewidth]{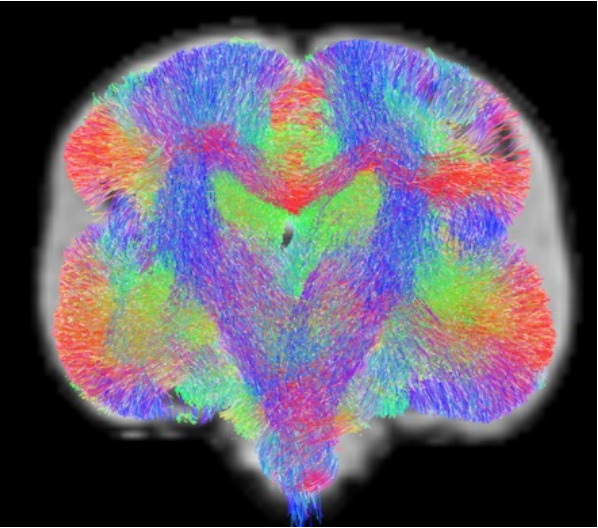} \\[2pt]

        \includegraphics[width=0.24\linewidth]{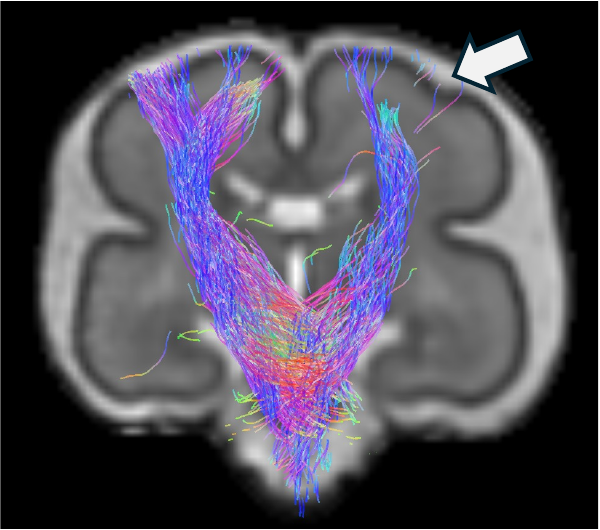} &
        \includegraphics[width=0.24\linewidth]{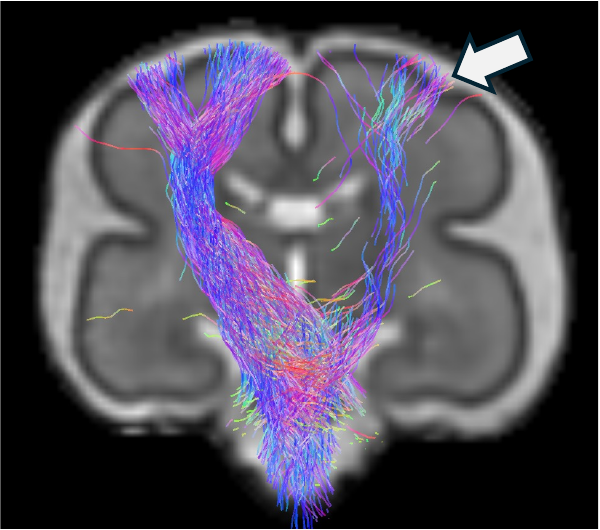} &
        \includegraphics[width=0.24\linewidth]{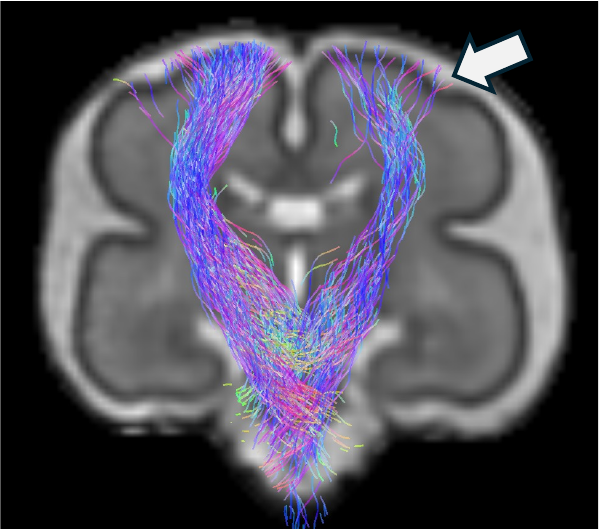} &
        \includegraphics[width=0.24\linewidth]{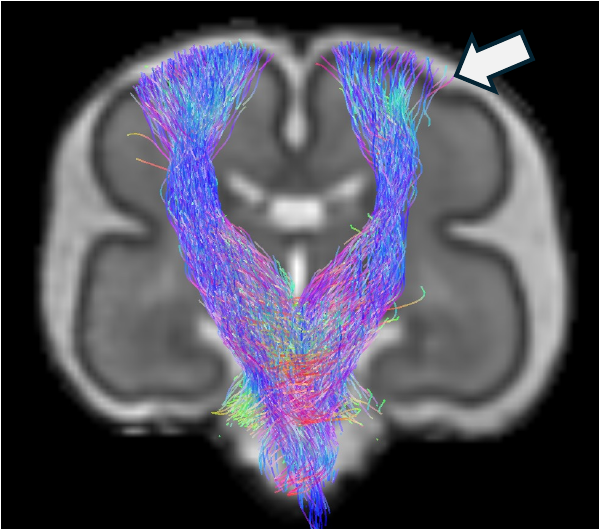} \\[2pt]

        \includegraphics[width=0.24\linewidth]{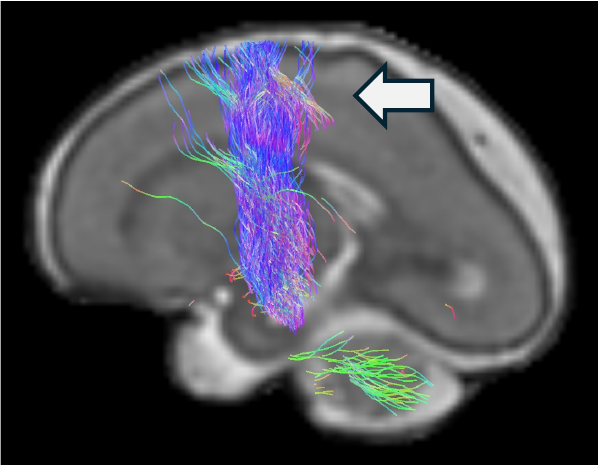} &
        \includegraphics[width=0.24\linewidth]{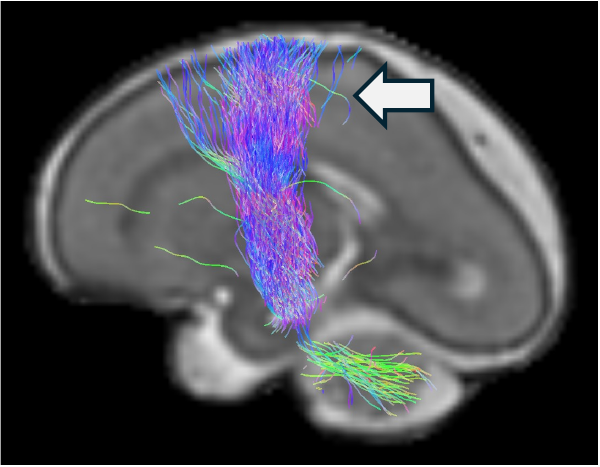} &
        \includegraphics[width=0.24\linewidth]{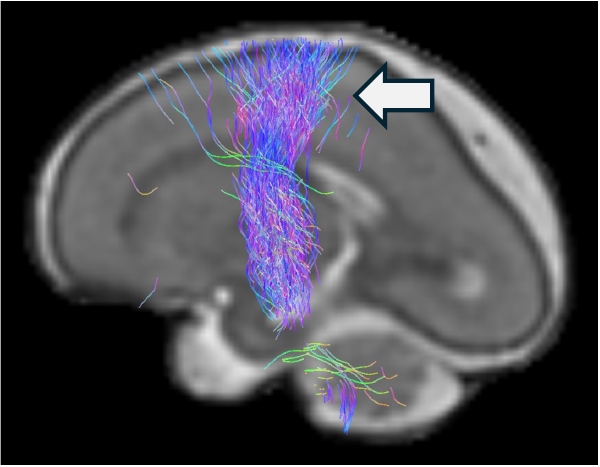} &
        \includegraphics[width=0.24\linewidth]{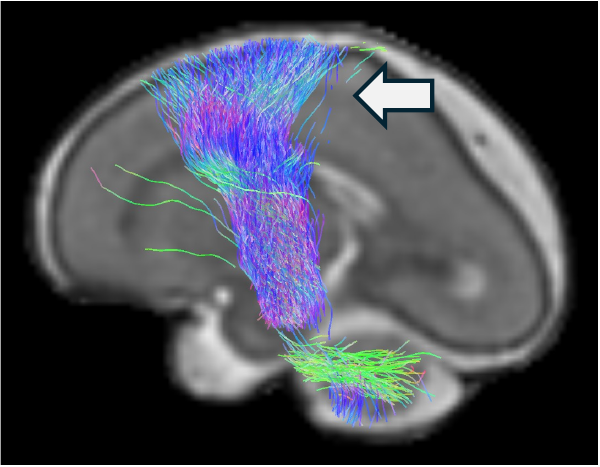}
    \end{tabular}

    \caption{Qualitative comparison of reconstructed tracts across methods on in-vivo fetal data. White arrows: cortico-spinal tract reaches the cortex more consistently and reliably in ANGEL compared to the other methods.}
    \label{fig:comparsion_fetal_tracts}
\end{figure}

%These quantitative gains are complemented by the qualitative comparisons.
Fig.~\ref{fig:comparsion_neonatal_dwi} shows reconstructed DWIs on
simulated neonatal data, where \textsc{Angel} produces closer anatomical agreement with the reference.
Fig.~\ref{fig:comparsion_fetal_tracts} compares tractography on
in-vivo fetal data. The reconstructed tracts from ANGEL form more coherent bundles, with example cortico-spinal tract reaching cortex in more reliable and anatomically plausible way. 
%with smoother trajectories and fewer apparent tracking artifacts.
%Together, these results support improved reconstruction in both the
%spatial and angular domains.

% \begin{figure}[!t]
% \centerline{\includegraphics[width=\columnwidth]{fig/vis_neonatal.pdf}}
% \caption{Qualitative comparison of DWI reconstructions across methods on simulated neonatal data.}
% \label{fig:comparsion_neonatal_dwi}
% \end{figure}

% \begin{figure}[!t]
% \centerline{\includegraphics[width=\columnwidth]{fig/comparison_vis.pdf}}
% \caption{Qualitative comparison of reconstructed tracts across methods on in vivo fetal data.}
% \label{fig:comparsion_fetal_tracts}
% \end{figure}

\begin{table}[t]
\caption{Quantitative comparison of DWI reconstruction on the fetal and simulated neonatal datasets. Values are mean$\pm$SD, with the best mean for each metric within each dataset in bold. Runtime is in minutes. $\dagger$ indicates significant improvement over all competing methods based on paired $t$-tests ($p<0.05$).}
\label{tab:main_dwi_comparison}
\centering
\scriptsize
\setlength{\tabcolsep}{3.5pt}
\begin{tabular}{llccc}
\toprule
Dataset & Method & PSNR $\uparrow$  &
NCC $\uparrow$  & Runtime $\downarrow$  \\
\midrule
\multirow{4}{*}{Fetal} & IRTK & $23.34\!\pm\!0.98$ & $0.8085\!\pm\!0.0350$  & $8.77\!\pm\!4.93$ \\
& SHINR & $23.54\!\pm\!1.23$ &
$0.8194\!\pm\!0.0374$  & $5.52\!\pm\!0.16$ \\
& Plain-GS & $23.86\!\pm\!1.81$ &
$0.8286\!\pm\!0.0583$ &  $\mathbf{3.00\!\pm\!0.04}$  \\
& \textsc{Angel} & $\mathbf{24.55\!\pm\!1.59}^{\dagger}$ & $\mathbf{0.8578\!\pm\!0.0401}^{\dagger}$  & $3.14\!\pm\!0.01$ \\
\midrule
\multirow{4}{*}{Neonatal} & IRTK & $24.64\!\pm\!0.57$  &
$0.8994\!\pm\!0.0191$ &  $35.16\!\pm\!2.44$  \\
& SHINR & $24.92\!\pm\!0.64$  &
$0.9028\!\pm\!0.0245$ &  $5.22\!\pm\!0.04$  \\
& Plain-GS & $25.07\!\pm\!0.52$  &
$0.9063\!\pm\!0.0155$  & $3.68\!\pm\!0.34$  \\
& \textsc{Angel} & $\mathbf{27.06\!\pm\!0.84}^{\dagger}$ &
  $\mathbf{0.9482\!\pm\!0.0120}^{\dagger}$ & $\mathbf{2.47\!\pm\!0.00}$  \\
\bottomrule
\end{tabular}
\end{table}

\begin{table*}[t]
  \centering
  \caption{Quantitative comparison of angular reconstruction on simulated neonatal data. Values are mean$\pm$SD across 20 subjects, with the best mean for each metric in bold. $\dagger$ indicates significant improvement over all competing methods based on paired $t$-tests ($p<0.05$).}
  \label{tab:main_angular_comparison}
  \setlength{\tabcolsep}{3.5pt}
  \renewcommand{\arraystretch}{1.08}
  \scriptsize
  \begin{tabular}{@{}lccccccc@{}}
    \toprule
    & \multicolumn{2}{c}{SH2} & \multicolumn{2}{c}{SH4}
    & \multicolumn{3}{c}{FOD} \\
    \cmidrule(lr){2-3}\cmidrule(lr){4-5}\cmidrule(lr){6-8}
    Method
    & PSNR $\uparrow$ & NCC $\uparrow$
    & PSNR $\uparrow$ & NCC $\uparrow$
    & Full ACC $\uparrow$ & Shape ACC $\uparrow$
    & PAE ($^{\circ}$) $\downarrow$ \\
    \midrule
    IRTK
    & $29.27 \pm 0.67$ & $0.7614 \pm 0.0287$
    & $21.58 \pm 1.36$ & $0.2952 \pm 0.0587$
    & $0.7324 \pm 0.0205$ & $0.4192 \pm 0.0391$
    & $33.54 \pm 1.16$ \\
    SHINR
    & $30.01 \pm 0.63$ & $0.8046 \pm 0.0271$
    & $24.29 \pm 0.81$ & $0.3167 \pm 0.0851$
    & $0.7753 \pm 0.0185$ & $0.4528 \pm 0.0412$
    & $32.97 \pm 1.29$ \\
    Plain-GS
    & $29.80 \pm 0.62$ & $0.7957 \pm 0.0384$
    & $21.28 \pm 1.71$ & $0.3488 \pm 0.0722$
    & $0.7383 \pm 0.0186$ & $0.4426 \pm 0.0408$
    & $33.13 \pm 1.21$ \\
    \textsc{Angel}
    & $\mathbf{32.85 \pm 0.68}^{\dagger}$ & $\mathbf{0.9098 \pm 0.0131}^{\dagger}$
    & $\mathbf{25.79 \pm 0.84}^{\dagger}$ & $\mathbf{0.5296 \pm 0.0590}^{\dagger}$
    & $\mathbf{0.8371 \pm 0.0145}^{\dagger}$ & $\mathbf{0.5856 \pm 0.0303}^{\dagger}$
    & $\mathbf{28.81 \pm 1.12}^{\dagger}$ \\
    \bottomrule
  \end{tabular}
\end{table*}

\subsection{Effectiveness of Cohort Training.}
To assess the effect of the cohort-training strategy described in Sec.~\ref{sec:cross_subject}, we evaluate angular reconstruction quantitatively on simulated neonatal data and qualitatively compare tractography on in-vivo fetal data. 

\noindent\textbf{Simulated neonatal data.}  
The 20 simulated neonatal subjects are divided into cohorts of $n\in{1,2,5,10,20}$ subjects. Each cohort was jointly reconstructed using a shared decoder.
For example, when $n=2$, the 20 subjects were randomly divided into ten two-subject cohorts, whereas $n=1$ corresponds to conventional subject-specific optimization. 

Table~\ref{tab:cohort_training} summarizes the resulting reconstruction performance.
Joint optimization of only two subjects already improved all reported angular metrics over the $n=1$ baseline, including the reconstruction of the second- and fourth-order SH components, the full and shape FOD ACC, and the peak angular error. Performance improved further on all metrics with larger cohorts before saturating, with minor metric-specific variations. In particular, the SH4 PSNR obtained with $n=20$ exceeded those obtained with $n=1, 2, 5, 10$ by 0.85, 0.47, 0.20, and 0.15 dB, respectively.

\begin{figure}[t]
    \centering
    \setlength{\tabcolsep}{1pt}
    \renewcommand{\arraystretch}{0}

    \begin{tabular}{@{}ccc@{}}
        \multicolumn{3}{c}{\scriptsize\textbf{Joint Cohort Training}} \\[2pt]

        \includegraphics[width=0.32\columnwidth]{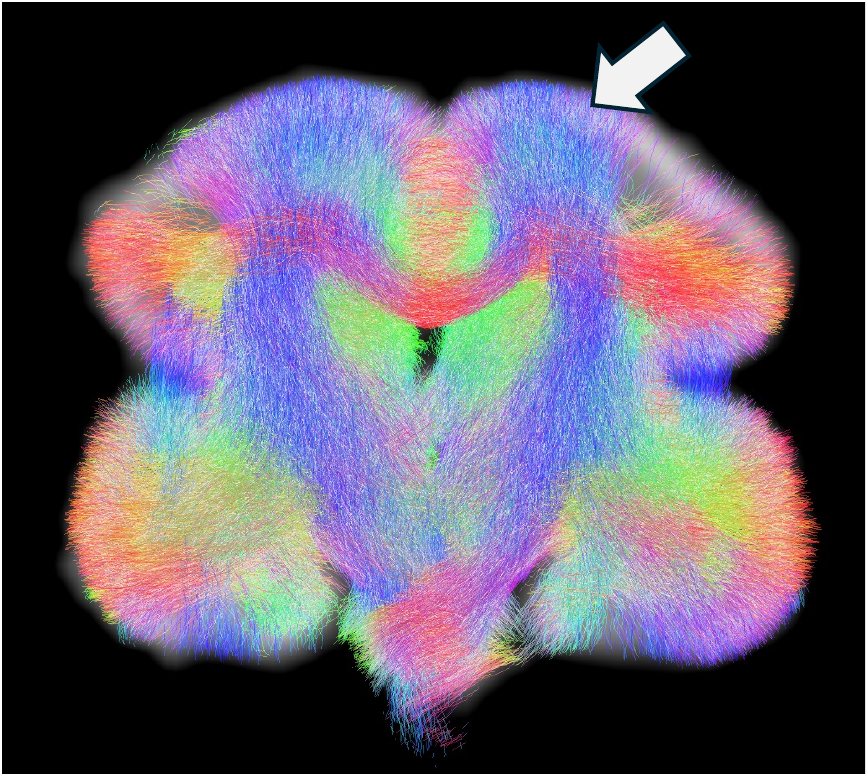} &
        \includegraphics[width=0.32\columnwidth]{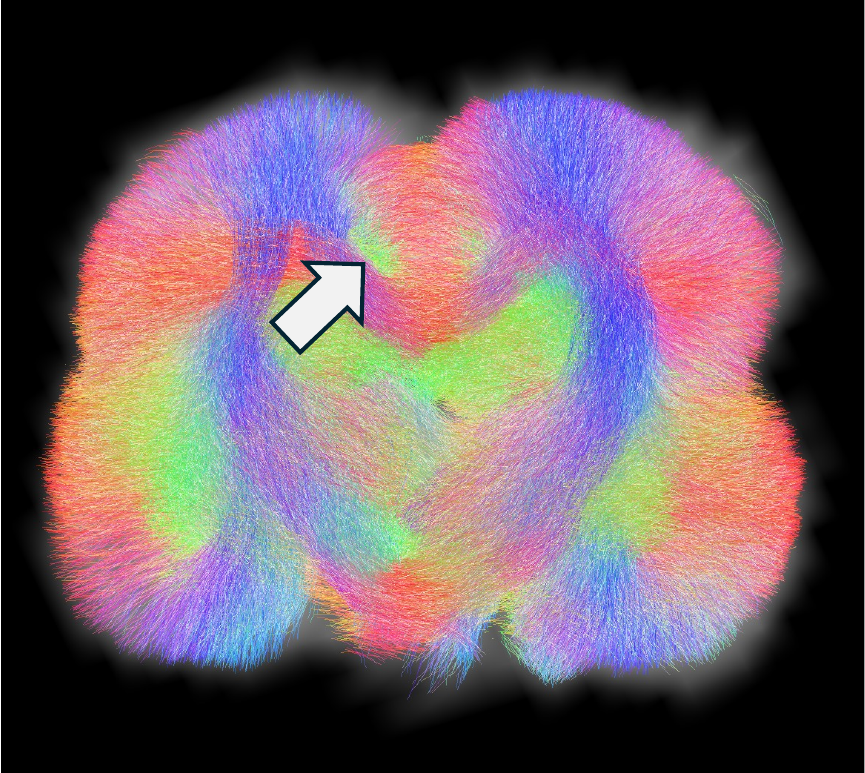} &
        \includegraphics[width=0.32\columnwidth]{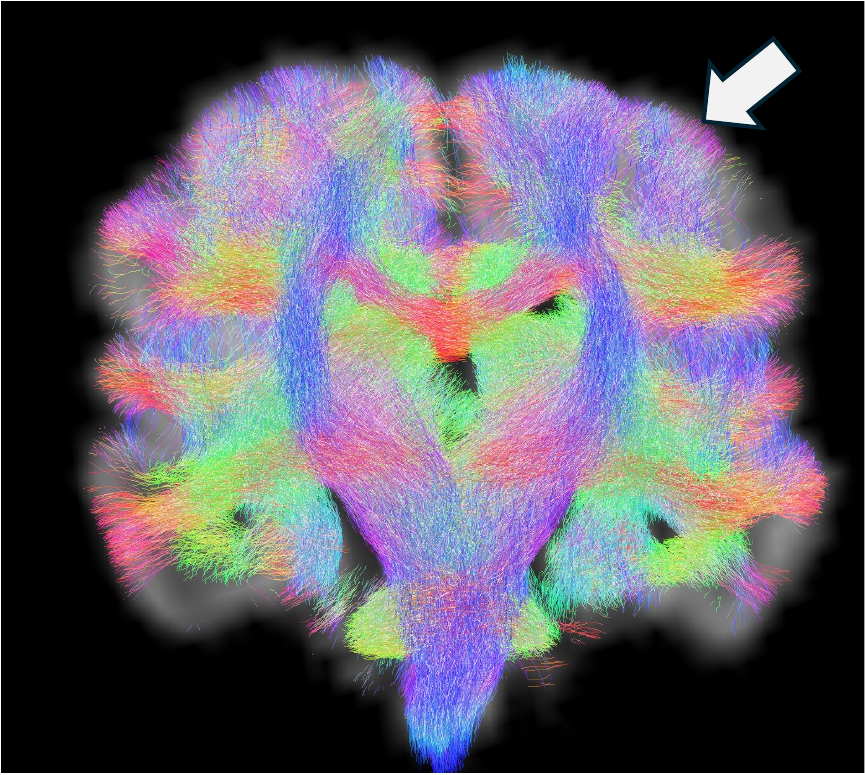} \\[4pt]

        \multicolumn{3}{c}{\scriptsize\textbf{Single-Subject Training}} \\[2pt]

        \includegraphics[width=0.32\columnwidth]{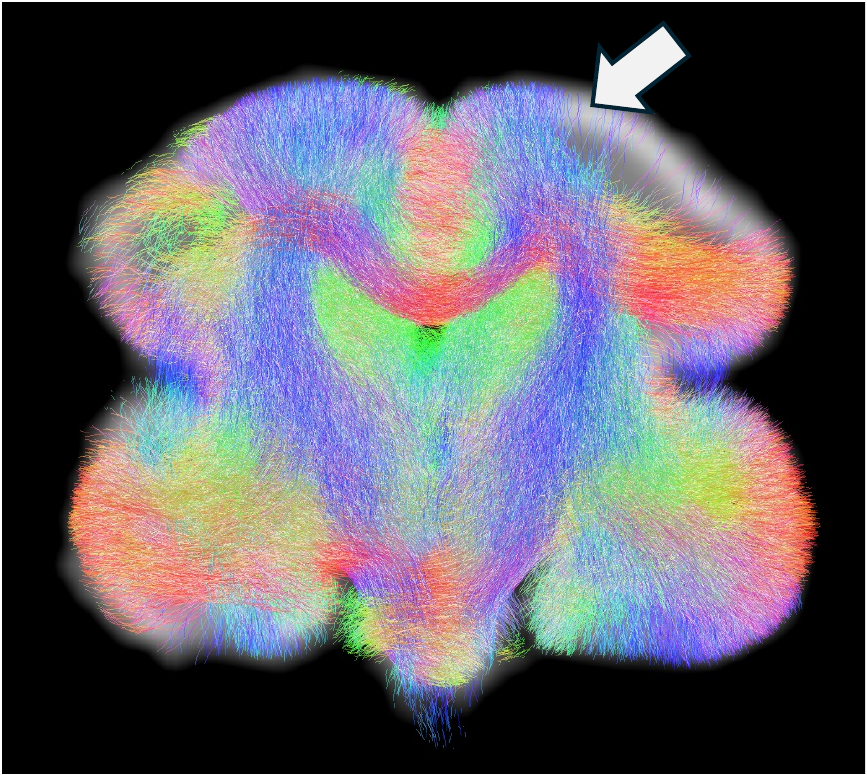} &
        \includegraphics[width=0.32\columnwidth]{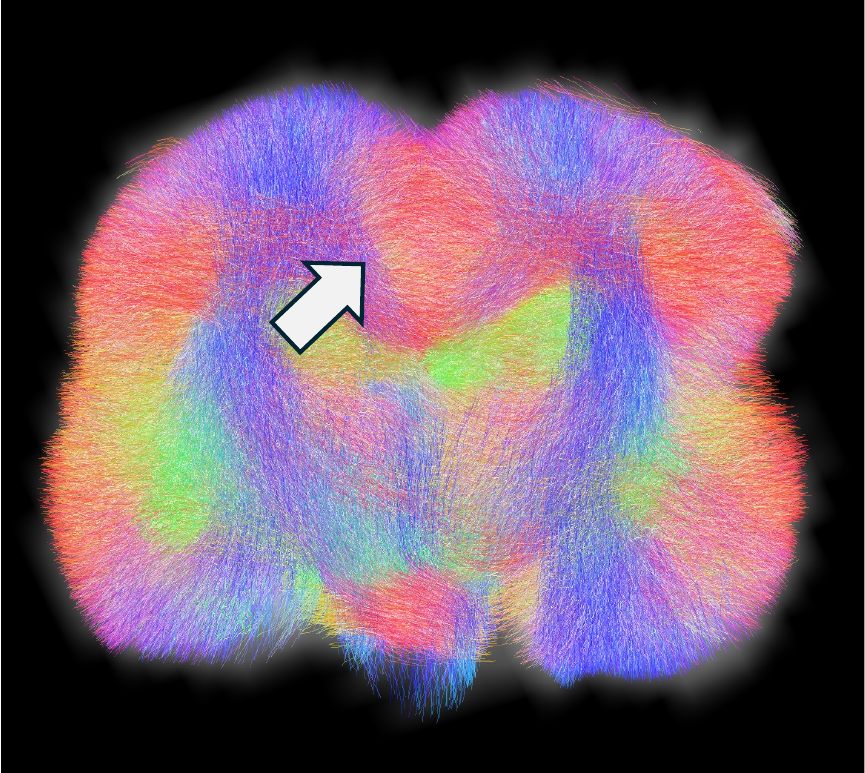} &
        \includegraphics[width=0.32\columnwidth]{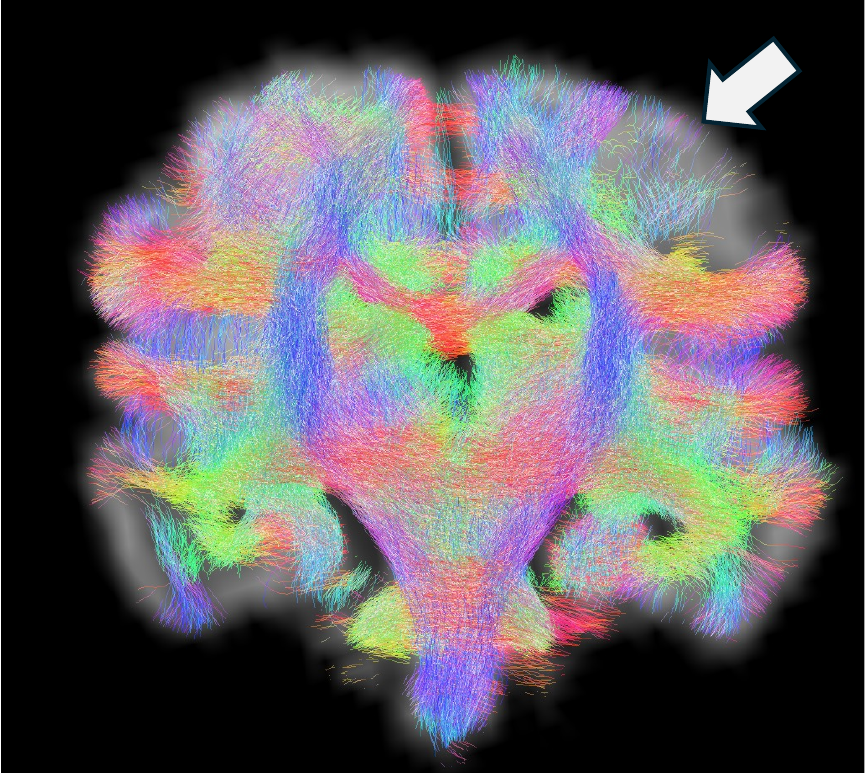}
    \end{tabular}

    \caption{Qualitative comparison of joint cohort training (top row) and single-subject training (bottom row) on three fetal subjects. White arrows: Corico-spinal tract (blue) reaches the cortex in joint training but not in single-subject training (left); Cingulum (green) is visible with joint cohort training but not in single subject training (middle); Streamlines are reaching the cortex more reliably in joint cohort training (right).}
    \label{fig:cohort_training}
\end{figure}

\noindent\textbf{In vivo fetal data.} Fig.~\ref{fig:cohort_training} illustrates the effect of cohort training on tractography in three fetal subjects. Compared with subject-specific optimization, joint cohort training yields less noisy streamlines, resulting in more reliable tracking (e.g. cortico-spinal tract reaching cortex in left and right subject) and better visualisation of the individual tract (e.g. cingulum in the middle subject). Together, these quantitative and qualitative results suggest that cross-subject decoder sharing improves angular reconstruction and helps limit noise overfitting in higher-order SH coefficients. 

% On simulated neonatal data, this benefit is already evident when only two subjects are jointly optimized, while larger cohorts generally provide further gains in angular fidelity without compromising DWI reconstruction.

% \begin{figure}[!t]
% \centerline{\includegraphics[width=0.9\columnwidth]{fig/ablation_cohort_vis.pdf}}
% \caption{Qualitative ablation of cohort training on three subjects. Each column shows tractography for one subject, with joint cohort training in the top row and single-subject training in the bottom row.}
% \label{fig:ablation_cohort}
% \end{figure}

\begin{table}[t]
  \centering
  \caption{Effect of cohort size on simulated neonatal reconstruction.
  Results are summarized across the same 20 subjects. Each cell reports
  the mean above and SD in parentheses below, with the best mean for
  each metric in bold. $\dagger$ indicates significant improvement over $n=1$ based on paired $t$-tests ($p<0.05$).}
  \label{tab:cohort_training}
  \scriptsize
  \setlength{\tabcolsep}{2.5pt}
  \renewcommand{\arraystretch}{1.05}
\newcommand{\cohortval}[3][]{%
  \begin{tabular}[c]{@{}c@{}}
    ${#2}^{#1}$ \\
    $(#3)$
  \end{tabular}}

\newcommand{\cohortbest}[3][]{%
  \begin{tabular}[c]{@{}c@{}}
    $\mathbf{#2}^{#1}$ \\
    $(#3)$
  \end{tabular}}
  \begin{tabular}{@{}llccccc@{}}
    \toprule
    & & \multicolumn{5}{c}{Cohort Size} \\
    \cmidrule(lr){3-7}
    Category & Metric & $n=1$ & $n=2$ & $n=5$ & $n=10$ & $n=20$ \\
    \midrule
    \multirow{2}{*}{DWI}
    & PSNR $\uparrow$
    & \cohortbest{27.18}{0.85} & \cohortval{27.17}{0.87}
    & \cohortbest{27.18}{0.82} & \cohortval{27.14}{0.85}
    & \cohortval{27.07}{0.84} \\
    & NCC $\uparrow$
    & \cohortval{0.9467}{0.0125} & \cohortval[\dagger]{0.9482}{0.0128}
    & \cohortbest[\dagger]{0.9493}{0.0120} & \cohortval[\dagger]{0.9491}{0.0121}
    & \cohortval{0.9483}{0.0120} \\
    \midrule
    \multirow{2}{*}{SH2}
    & PSNR $\uparrow$
    & \cohortval{32.01}{0.56} & \cohortval[\dagger]{32.53}{0.61}
    & \cohortval[\dagger]{32.88}{0.66} & \cohortbest[\dagger]{32.89}{0.67}
    & \cohortval[\dagger]{32.85}{0.66} \\
    & NCC $\uparrow$
    & \cohortval{0.8817}{0.0143} & \cohortval[\dagger]{0.8968}{0.0122}
    & \cohortval[\dagger]{0.9065}{0.0118} & \cohortval[\dagger]{0.9081}{0.0130}
    & \cohortbest[\dagger]{0.9098}{0.0128} \\
    \midrule
    \multirow{2}{*}{SH4}
    & PSNR $\uparrow$
    & \cohortval{24.94}{1.18} & \cohortval[\dagger]{25.33}{1.02}
    & \cohortval[\dagger]{25.60}{0.92} & \cohortval[\dagger]{25.64}{0.89}
    & \cohortbest[\dagger]{25.79}{0.82} \\
    & NCC $\uparrow$
    & \cohortval{0.4822}{0.0801} & \cohortval[\dagger]{0.5097}{0.0703}
    & \cohortval[\dagger]{0.5295}{0.0658} & \cohortval[\dagger]{0.5266}{0.0617}
    & \cohortbest[\dagger]{0.5296}{0.0576} \\
    \midrule
    \multirow{3}{*}{FOD}
    & PAE ($^\circ$) $\downarrow$
    & \cohortval{30.09}{1.26} & \cohortval[\dagger]{29.34}{1.22}
    & \cohortbest[\dagger]{28.74}{1.22} & \cohortval[\dagger]{28.85}{1.18}
    & \cohortval[\dagger]{28.81}{1.09} \\
    & Full ACC $\uparrow$
    & \cohortval{0.8131}{0.0154} & \cohortval[\dagger]{0.8250}{0.0132}
    & \cohortval[\dagger]{0.8329}{0.0141} & \cohortval[\dagger]{0.8336}{0.0134}
    & \cohortbest[\dagger]{0.8371}{0.0141} \\
    & Shape ACC $\uparrow$
    & \cohortval{0.5469}{0.0379} & \cohortval[\dagger]{0.5699}{0.0336}
    & \cohortbest[\dagger]{0.5863}{0.0342} & \cohortval[\dagger]{0.5831}{0.0313}
    & \cohortval[\dagger]{0.5856}{0.0295} \\
    \bottomrule
  \end{tabular}
\end{table}

\subsection{Effectiveness of Unseen-Subject Adaptation}
To evaluate the unseen-subject adaptation described in Sec.~\ref{sec:cross_subject}, the 20 neonatal subjects were split into disjoint pretraining and held-out sets of 10 subjects each. Separate decoders $\Theta$ were pretrained using $n\in{1,2,5,10}$ subjects, with the $n<10$ subsets randomly sampled from the fixed pretraining set. For adaptation, the pretrained decoder was frozen, while a new subject-specific scaffold and the remaining acquisition parameters were optimized. The Scratch baseline instead optimized the complete representation independently for each held-out subject without decoder pretraining.

\noindent\textbf{Effect of pre-training.} Table~\ref{tab:heldout_adaptation} summarizes the adaptation results. Decoder pretraining improved the DWI metrics for every pretraining cohort size, demonstrating that the learned representation transfers beyond the subjects used for pretraining. Angular reconstruction was more dependent on cohort diversity: pretraining on a single subject improved the SH2 metrics but slightly degraded the SH4 and FOD measures, suggesting that a decoder constrained by only one subject may retain subject-specific high-order variations. This limitation diminished as the pretraining cohort increased. With at least two subjects, pretraining achieved better performance than Scratch on all metrics, and $n=10$ achieved the best performance overall. These results show that sufficient cohort diversity enables the shared decoder to serve as a transferable spatial-angular prior, improving noise-sensitive high-order angular reconstruction without compromising DWI fidelity.

\begin{figure}[!t]
\centerline{\includegraphics[width=\columnwidth]{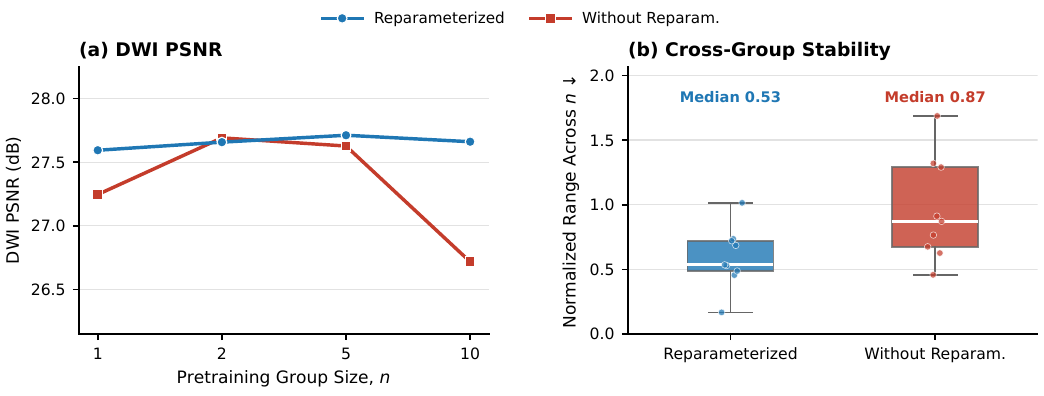}}
\caption{Effect of latent reparameterization on adaptation across pretraining cohort sizes. (a) Mean DWI PSNR over 10 held-out subjects shows better performance and stability with different cohort sizes with latent reparametrisation. (b) Performance range across pretraining cohort sizes for 9 metrics, normalized by the median subject-level standard deviation. Each point represents one metric; lower values indicate greater stability.}
\label{fig:repar}
\end{figure}

\noindent\textbf{Effect of Latent Reparameterization.} Fig.~\ref{fig:repar} evaluates the latent reparameterization introduced in Eq.~\ref{eq:repa} across models pretrained with different cohort sizes \(n\). The subject-specific adaptation is performed using the same optimization configuration for both variants (with/without reparametrisation) across all values of \(n\). As shown in Fig.~\ref{fig:repar} (a), reparameterization maintains stable DWI PSNR across cohort sizes, whereas direct optimization of the scaffold degrades at the smallest and largest \(n\), indicating its sensitivity to changes in latent feature dispersion. Fig.~\ref{fig:repar} (b) extends this analysis to nine metrics. For each metric, the mean and standard deviation are computed over the same 10 held-out subjects at each pretraining cohort size $n \in {1, 2, 5, 10}$, separately with and without reparameterization. The normalized range is calculated by dividing the difference between the largest and smallest of the four means by the median of the four corresponding standard deviations. With reparameterization, the median normalized range across the nine metrics decreases from 0.87 to 0.53, a 39\% reduction. This shows that the proposed reparameterization improves adaptation stability across different pretraining cohort sizes and avoids cohort-specific tuning.

\begin{table}[t]
  \centering
  \caption{Held-out adaptation with a frozen pretrained decoder. Each cell reports the mean above and SD in parentheses below across
  10 held-out subjects, with the best mean in bold.
  $\dagger$ indicates significant improvement over Scratch based on
  paired t-test ($p<0.05$).}
  \label{tab:heldout_adaptation}
  \scriptsize
  \setlength{\tabcolsep}{2.5pt}
  \renewcommand{\arraystretch}{1.05}
  \newcommand{\heldoutval}[3][]{%
    \begin{tabular}[c]{@{}c@{}}
      $#2^{#1}$ \\ $(#3)$
    \end{tabular}}
  \newcommand{\heldoutbest}[3][]{%
    \begin{tabular}[c]{@{}c@{}}
      $\mathbf{#2}^{#1}$ \\ $(#3)$
    \end{tabular}}
  \begin{tabular}{@{}llccccc@{}}
    \toprule
    & & & \multicolumn{4}{c}{Pretraining Cohort Size} \\
    \cmidrule(lr){4-7}
    Category & Metric & Scratch & $n=1$ & $n=2$ & $n=5$ & $n=10$ \\
    \midrule
    \multirow{2}{*}{DWI}
    & PSNR $\uparrow$
    & \heldoutval{27.15}{0.72} & \heldoutval[\dagger]{27.60}{0.67}
    & \heldoutval[\dagger]{27.66}{0.66}
    & \heldoutbest[\dagger]{27.70}{0.67}
    & \heldoutval[\dagger]{27.66}{0.69} \\
    & NCC $\uparrow$
    & \heldoutval{0.9468}{0.0053} & \heldoutval[\dagger]{0.9499}{0.0060}
    & \heldoutval[\dagger]{0.9512}{0.0055}
    & \heldoutval[\dagger]{0.9520}{0.0054}
    & \heldoutbest[\dagger]{0.9528}{0.0053} \\
    \midrule
    \multirow{2}{*}{SH2}
    & PSNR $\uparrow$
    & \heldoutval{31.87}{0.70} & \heldoutval[\dagger]{32.09}{0.78}
    & \heldoutval[\dagger]{32.26}{0.79}
    & \heldoutval[\dagger]{32.33}{0.77}
    & \heldoutbest[\dagger]{32.47}{0.83} \\
    & NCC $\uparrow$
    & \heldoutval{0.8829}{0.0176} & \heldoutval[\dagger]{0.8881}{0.0164}
    & \heldoutval[\dagger]{0.8928}{0.0167}
    & \heldoutval[\dagger]{0.8972}{0.0157}
    & \heldoutbest[\dagger]{0.8998}{0.0161} \\
    \midrule
    \multirow{2}{*}{SH4}
    & PSNR $\uparrow$
    & \heldoutval{25.26}{1.30} & \heldoutval{25.11}{1.14}
    & \heldoutval{25.43}{1.10} & \heldoutval[\dagger]{25.54}{1.13}
    & \heldoutbest[\dagger]{25.69}{1.03} \\
    & NCC $\uparrow$
    & \heldoutval{0.5021}{0.0821} & \heldoutval{0.4756}{0.0707}
    & \heldoutval{0.5033}{0.0703} & \heldoutval{0.5142}{0.0729}
    & \heldoutbest{0.5153}{0.0660} \\
    \midrule
    \multirow{3}{*}{FOD}
    & PAE ($^\circ$) $\downarrow$
    & \heldoutval{29.68}{1.30} & \heldoutval{30.00}{1.28}
    & \heldoutval[\dagger]{29.29}{1.29} & \heldoutval[\dagger]{29.18}{1.31}
    & \heldoutbest[\dagger]{29.02}{1.10} \\
    & Full ACC $\uparrow$
    & \heldoutval{0.8184}{0.0154} & \heldoutval{0.8139}{0.0129}
    & \heldoutval[\dagger]{0.8234}{0.0148} & \heldoutval[\dagger]{0.8263}{0.0151}
    & \heldoutbest[\dagger]{0.8295}{0.0144} \\
    & Shape ACC $\uparrow$
    & \heldoutval{0.5633}{0.0384} & \heldoutval{0.5521}{0.0351}
    & \heldoutval{0.5721}{0.0363} & \heldoutval[\dagger]{0.5779}{0.0365}
    & \heldoutbest[\dagger]{0.5797}{0.0322} \\
    \bottomrule
  \end{tabular}
\end{table}

\subsection{Effectiveness of Bias-Field Correction.}
We evaluate six progressively stronger levels of slice-dependent bias
corruption simulated using MONAI~\cite{monai}, comparing \textsc{Angel},
\textit{Plain-GS}, and SHINR with and without bias field correction across nine
spatial and angular metrics. Both Gaussian methods employ our
slice-conditioned bias representation. Its primitives share signal
anchors and rendering neighborhoods in \textsc{Angel}, but are positioned
independently in \textit{Plain-GS}. SHINR uses its native INR-based model.

\begin{figure}[!t]
\centerline{\includegraphics[width=\columnwidth]{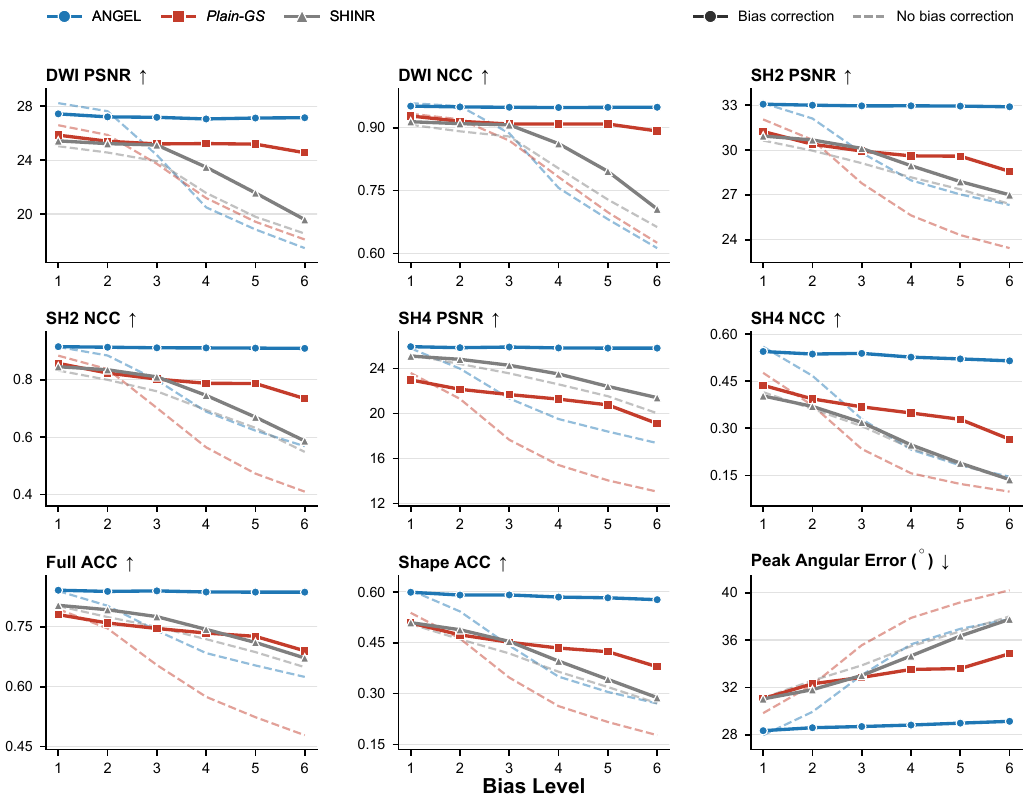}}
\caption{Robustness to bias-field corruption across six bias levels. Curves show mean performance over 20 subjects. Solid and dashed lines denote results with and without bias correction, respectively.}
\label{fig:snr_sweep}
\end{figure}

As shown in Fig.~\ref{fig:snr_sweep}, \textsc{Angel} remains stable
across all six levels, with DWI PSNR decreasing by only 0.29~dB and peak
angular error increasing by $0.79^\circ$. Although bias field correction offers
limited benefit under weak corruption, it improves all nine metrics
for both Gaussian methods from level 3 onward. At the strongest level,
enabling bias field correction improves DWI PSNR by 9.68~dB in \textsc{Angel} and
6.45~dB in \textit{Plain-GS}, while reducing peak angular error by
$8.63^\circ$ and $5.36^\circ$, respectively. SHINR degrades more substantially despite its native correction.
At level 6, \textsc{Angel} achieves 7.57~dB higher DWI PSNR and 0.379
higher SH4 NCC than SHINR. These results demonstrate the effectiveness
of the proposed bias module in both Gaussian representations, with
\textsc{Angel} providing the strongest overall reconstruction under
severe intensity corruption.

% \subsection{Noise Robustness under Varying SNR}

% Robustness to measurement noise was evaluated at five input SNR levels ranging from 20 to 40, with SNR defined relative to the $b=0$ images. As shown in Fig.~\ref{fig:snr_sweep}, the proposed method consistently outperformed Plain-GS and SHINR across all nine metrics covering DWI fidelity, SH reconstruction, FOD accuracy, and peak orientation accuracy. The advantage became particularly pronounced under severe noise. At SNR 20, the proposed method remained superior to both baselines even when they were evaluated at SNR 40. At this lowest SNR, DWI PSNR was improved by 2.38 and 1.93 dB over Plain-GS and SHINR, respectively, while the two-fiber peak angular error was reduced by 3.29$^\circ$ and 3.61$^\circ$. The gains were also more pronounced for SH4 than SH2 and for Shape ACC than Full ACC, indicating that the proposed representation particularly improves the recovery of higher-order angular structure rather than merely enhancing overall signal fidelity. These results demonstrate strong robustness of the proposed spatial-angular representation to low-SNR diffusion measurements.

% Fig. 3: Representative ground truth, corrupted input, reconstructions, and
% matched error maps in axial/coronal/sagittal views.
\section{Conclusion}
This paper presents \textsc{Angel}, an anchor-structured neural Gaussian framework for fetal dMRI SVR with cross-subject learning. Our experiments demonstrate that proposed Gaussian representation based on subject-specific anchor features and cross-subject cross-anchor decoder significantly outperforms baseline Gaussian representation, as well as implicit neural representations and classical super-resolution reconstruction, in both spatial and angular domains. Our framework retains flexibility and efficiency of the Gaussian representation, while reduced parameter space and spatial-angular signal regularisation learned from multiple subjects prevents overfitting to noise and image artefacts, resulting in better consistency with acquired dMRI data. Our qualitative results show more accurate and reliable tractography, demonstrating positive impact on downstream application. 

%through a neural decoder shared across anchors and subjects. This shared parameterization allows cohort observations to regularize higher-order angular estimation and limit noise overfitting under low-SNR conditions, while subject-specific anchors retain individual anatomy. The same anchors provide shared spatial support for the diffusion signal and bias field, while differences in Gaussian primitive count and scale allow smooth, slice-dependent intensity modulation to be modeled separately from the underlying diffusion signal. Experiments on simulated neonatal and in vivo fetal data demonstrate improved spatial and angular reconstruction over existing methods.

\section{LLM Usage Statement}

{
\emergencystretch=2em
\looseness=-1
ChatGPT (GPT-6.1 Sol) was used for manuscript editing, while Claude Opus 5.5 assisted with coding. Both models supported discussions on methodological and experimental choices. All AI-assisted outputs were reviewed by the authors, who take full responsibility for the work.\par
}

\section*{Acknowledgments}
This research was supported by the European Research
Council under the European Union’s Seventh Framework Programme [FP7/ 20072013]/ERC grant
agreement no. 319456 Dhcp project and the National Institute for Health and Care Research (NIHR) Clinical Research Facility at Guy’s and St Thomas’ NHS Foundation Trust. The views expressed are those of the author(s) and not necessarily those of the NHS, the NIHR or the Department of Health and Social Care. We also thank Jun Wang for suggesting the lovely name
\textsc{Angel} for our method.
{
    \small
    \bibliographystyle{ieeenat_fullname}
    \bibliography{main}

@String(CVPR  = {IEEE Conf. Comput. Vis. Pattern Recog.})

@String(AAAI  = {AAAI})

@String(CVPR  = {CVPR})

@article{svrtk,
  title={Reconstruction of fetal brain MRI with intensity matching and complete outlier removal},
  author={Kuklisova-Murgasova, Maria and Quaghebeur, Gerardine and Rutherford, Mary A and Hajnal, Joseph V and Schnabel, Julia A},
  journal={Medical image analysis},
  volume={16},
  number={8},
  pages={1550--1564},
  year={2012},
  publisher={Elsevier}
}

@article{nesvor,
  title={NeSVoR: implicit neural representation for slice-to-volume reconstruction in MRI},
  author={Xu, Junshen and Moyer, Daniel and Gagoski, Borjan and Iglesias, Juan Eugenio and Grant, P Ellen and Golland, Polina and Adalsteinsson, Elfar},
  journal={IEEE transactions on medical imaging},
  volume={42},
  number={6},
  pages={1707--1719},
  year={2023},
  publisher={IEEE}
}

@article{svr3-label-free,
  title={SAFFRON: A Physics-informed, Label-free Self-supervised Deep Learning Framework for Fast and Accurate 3D Fetal Brain MRI Reconstruction},
  author={Wu, Jiangjie and Sun, Taotao and Wang, Lihui and Zhang, Yuyao and Wei, Hongjiang},
  journal={IEEE Journal of Selected Topics in Signal Processing},
  year={2025},
  publisher={IEEE}
}

@inproceedings{svr4-mamba,
  title={SVRMamba: Slice-to-Volume Reconstruction from Multiple MRI Stacks with Slice Sequence Guided Mamba},
  author={Wu, Jiangjie and Wei, Hongjiang and Zhang, Yuyao},
  booktitle={Proceedings of the AAAI Conference on Artificial Intelligence},
  volume={39},
  number={8},
  pages={8404--8412},
  year={2025}
}

@inproceedings{svr-gaussian1,
  title={Self-Supervised Slice-to-Volume Reconstruction with Gaussian Representations for Fetal MRI},
  author={Wang, Yinsong and Fletcher, Thomas and Luo, Xinzhe and Dineen, {\'A}ine Travers and Cusack, Rhodri and Qin, Chen},
  booktitle={2026 IEEE 23rd International Symposium on Biomedical Imaging (ISBI)},
  pages={01--05},
  year={2026},
  organization={IEEE}
}

@article{svr-gaussian2,
  title={M-Gaussian: An Magnetic Gaussian Framework for Efficient Multi-Stack MRI Reconstruction},
  author={Zheng, Kangyuan and Cai, Xuan and Wang, Jiangqi and Fu, Guixing and Li, Zhuoshuo and Chen, Yazhou and Ge, Xinting and Qu, Liangqiong and Liu, Mengting},
  journal={arXiv preprint arXiv:2603.00145},
  year={2026}
}

@inproceedings{svort,
  title={SVoRT: Iterative transformer for slice-to-volume registration in fetal brain MRI},
  author={Xu, Junshen and Moyer, Daniel and Grant, P Ellen and Golland, Polina and Iglesias, Juan Eugenio and Adalsteinsson, Elfar},
  booktitle={International Conference on Medical Image Computing and Computer-Assisted Intervention},
  pages={3--13},
  year={2022},
  organization={Springer}
}

@article{diffusion-mri-recon,
  title={Higher order spherical harmonics reconstruction of fetal diffusion MRI with intensity correction},
  author={Deprez, Maria and Price, Anthony and Christiaens, Daan and Estrin, Georgia Lockwood and Cordero-Grande, Lucilio and Hutter, Jana and Daducci, Alessandro and Tournier, Jacques-Donald and Rutherford, Mary and Counsell, Serena J and others},
  journal={IEEE transactions on medical imaging},
  volume={39},
  number={4},
  pages={1104--1113},
  year={2020},
  publisher={IEEE}
}

@article{classic-1,
  title={Registration-based approach for reconstruction of high-resolution in utero fetal MR brain images},
  author={Rousseau, Francois and Glenn, Orit A and Iordanova, Bistra and Rodriguez-Carranza, Claudia and Vigneron, Daniel B and Barkovich, James A and Studholme, Colin},
  journal={Academic radiology},
  volume={13},
  number={9},
  pages={1072--1081},
  year={2006},
  publisher={Elsevier}
}

@article{classic-2,
  title={Robust super-resolution volume reconstruction from slice acquisitions: application to fetal brain MRI},
  author={Gholipour, Ali and Estroff, Judy A and Warfield, Simon K},
  journal={IEEE transactions on medical imaging},
  volume={29},
  number={10},
  pages={1739--1758},
  year={2010},
  publisher={IEEE}
}

@inproceedings{intro_deep1,
  title={Deep residual learning for image recognition},
  author={He, Kaiming and Zhang, Xiangyu and Ren, Shaoqing and Sun, Jian},
  booktitle={Proc. IEEE Conf. Comput. Vis. Pattern Recognit. (CVPR)},
  pages={770--778},
  year={2016}
}

@article{intro_deep2,
  title={ImageNet classification with deep convolutional neural networks},
  author={Krizhevsky, Alex and Sutskever, Ilya and Hinton, Geoffrey E},
  journal={Commun. ACM},
  volume={60},
  number={6},
  pages={84--90},
  year={2017},
  publisher={AcM New York, NY, USA}
}

@article{intro_deep3,
  title={Backpropagation applied to handwritten zip code recognition},
  author={LeCun, Yann and Boser, Bernhard and Denker, John S and Henderson, Donnie and Howard, Richard E and Hubbard, Wayne and Jackel, Lawrence D},
  journal={Neural computation},
  volume={1},
  number={4},
  pages={541--551},
  year={1989},
  publisher={MIT Press}
}

@article{intro_deep4,
  title={Attention is all you need},
  author={Vaswani, Ashish and Shazeer, Noam and Parmar, Niki and Uszkoreit, Jakob and Jones, Llion and Gomez, Aidan N and Kaiser, {\L}ukasz and Polosukhin, Illia},
  journal={Proc. Adv. Neural Inf. Process. Syst.},
  volume={30},
  year={2017}
}

@article{inr1,
  title={Nerf: Representing scenes as neural radiance fields for view synthesis},
  author={Mildenhall, Ben and Srinivasan, Pratul P and Tancik, Matthew and Barron, Jonathan T and Ramamoorthi, Ravi and Ng, Ren},
  journal={Communications of the ACM},
  volume={65},
  number={1},
  pages={99--106},
  year={2021},
  publisher={ACM New York, NY, USA}
}

@article{inr2,
  title={Scene representation networks: Continuous 3d-structure-aware neural scene representations},
  author={Sitzmann, Vincent and Zollh{\"o}fer, Michael and Wetzstein, Gordon},
  journal={Advances in neural information processing systems},
  volume={32},
  year={2019}
}

@article{dhcp1-diffusion,
  title={Time-efficient and flexible design of optimized multishell HARDI diffusion},
  author={Hutter, Jana and Tournier, J Donald and Price, Anthony N and Cordero-Grande, Lucilio and Hughes, Emer J and Malik, Shaihan and Steinweg, Johannes and Bastiani, Matteo and Sotiropoulos, Stamatios N and Jbabdi, Saad and others},
  journal={Magnetic resonance in medicine},
  volume={79},
  number={3},
  pages={1276--1292},
  year={2018},
  publisher={Wiley Online Library}
}

@article{dhcp2-diffusion,
  title={A data-driven approach to optimising the encoding for multi-shell diffusion MRI with application to neonatal imaging},
  author={Tournier, Jacques-Donald and Christiaens, Daan and Hutter, Jana and Price, Anthony N and Cordero-Grande, Lucilio and Hughes, Emer and Bastiani, Matteo and Sotiropoulos, Stamatios N and Smith, Stephen M and Rueckert, Daniel and others},
  journal={NMR in Biomedicine},
  volume={33},
  number={9},
  pages={e4348},
  year={2020},
  publisher={Wiley Online Library}
}

@article{diffusion-mri-recon-shard,
  title={Scattered slice SHARD reconstruction for motion correction in multi-shell diffusion MRI},
  author={Christiaens, Daan and Cordero-Grande, Lucilio and Pietsch, Maximilian and Hutter, Jana and Price, Anthony N and Hughes, Emer J and Vecchiato, Katy and Deprez, Maria and Edwards, A David and Hajnal, Joseph V and others},
  journal={Neuroimage},
  volume={225},
  pages={117437},
  year={2021},
  publisher={Elsevier}
}

@article{csd1,
  title={Direct estimation of the fiber orientation density function from diffusion-weighted MRI data using spherical deconvolution},
  author={Tournier, J-Donald and Calamante, Fernando and Gadian, David G and Connelly, Alan},
  journal={Neuroimage},
  volume={23},
  number={3},
  pages={1176--1185},
  year={2004},
  publisher={Elsevier}
}

@article{csd2,
  title={Robust determination of the fibre orientation distribution in diffusion MRI: non-negativity constrained super-resolved spherical deconvolution},
  author={Tournier, J-Donald and Calamante, Fernando and Connelly, Alan},
  journal={Neuroimage},
  volume={35},
  number={4},
  pages={1459--1472},
  year={2007},
  publisher={Elsevier}
}

@article{csd3,
  title={Multi-tissue constrained spherical deconvolution for improved analysis of multi-shell diffusion MRI data},
  author={Jeurissen, Ben and Tournier, Jacques-Donald and Dhollander, Thijs and Connelly, Alan and Sijbers, Jan},
  journal={NeuroImage},
  volume={103},
  pages={411--426},
  year={2014},
  publisher={Elsevier}
}

@article{intro-diffusion-mri1,
  title={In utero tractography of fetal white matter development},
  author={Kasprian, Gregor and Brugger, Peter C and Weber, Michael and Krss{\'a}k, Martin and Krampl, Elisabeth and Herold, Christian and Prayer, Daniela},
  journal={Neuroimage},
  volume={43},
  number={2},
  pages={213--224},
  year={2008},
  publisher={Elsevier}
}

@article{intro-diffusion-mri2,
  title={White matter maturation of normal human fetal brain. An in vivo diffusion tensor tractography study},
  author={Zanin, Emilie and Ranjeva, Jean-Philippe and Confort-Gouny, Sylviane and Guye, Maxime and Denis, Daniele and Cozzone, Patrick J and Girard, Nadine},
  journal={Brain and behavior},
  volume={1},
  number={2},
  pages={95--108},
  year={2011},
  publisher={Wiley Online Library}
}

@article{intro-diffusion-mri3,
  title={Assessing prenatal white matter connectivity in commissural agenesis},
  author={Kasprian, Gregor and Brugger, Peter C and Sch{\"o}pf, Veronika and Mitter, Christian and Weber, Michael and Hainfellner, Johannes A and Prayer, Daniela},
  journal={Brain},
  volume={136},
  number={1},
  pages={168--179},
  year={2013},
  publisher={Oxford University Press}
}

@article{intro-diffusion-mri4,
  title={Probabilistic tractography using Q-ball imaging and particle filtering: application to adult and in-utero fetal brain studies},
  author={Pontabry, Julien and Rousseau, Fran{\c{c}}ois and Oubel, Estanislao and Studholme, Colin and Koob, M{\'e}riam and Dietemann, J-L},
  journal={Medical image analysis},
  volume={17},
  number={3},
  pages={297--310},
  year={2013},
  publisher={Elsevier}
}

@article{intro-diffusion-mri5,
  title={Disrupted developmental organization of the structural connectome in fetuses with corpus callosum agenesis},
  author={Jakab, Andr{\'a}s and Kasprian, Gregor and Schwartz, Ernst and Gruber, Gerlinde Maria and Mitter, Christian and Prayer, Daniela and Sch{\"o}pf, Veronika and Langs, Georg},
  journal={Neuroimage},
  volume={111},
  pages={277--288},
  year={2015},
  publisher={Elsevier}
}

@article{intro-diffusion-mri6,
  title={In vivo tractography of fetal association fibers},
  author={Mitter, Christian and Prayer, Daniela and Brugger, Peter C and Weber, Michael and Kasprian, Gregor},
  journal={PloS one},
  volume={10},
  number={3},
  pages={e0119536},
  year={2015},
  publisher={Public Library of Science San Francisco, CA USA}
}

@article{dhcp2,
  title={Complex diffusion-weighted image estimation via matrix recovery under general noise models},
  author={Cordero-Grande, Lucilio and Christiaens, Daan and Hutter, Jana and Price, Anthony N and Hajnal, Jo V},
  journal={Neuroimage},
  volume={200},
  pages={391--404},
  year={2019},
  publisher={Elsevier}
}

@inproceedings{svr-gaussian-fast,
  title={Fast and Explicit: Slice-to-Volume Reconstruction via 3D Gaussian Primitives with Analytic Point Spread Function Modeling},
  author={Dannecker, Maik and Jia, Steven and Stolt-Ans{\'o}, Nil and Girard, Nadine and Auzias, Guillaume and Rousseau, Fran{\c{c}}ois and Rueckert, Daniel},
  booktitle={Medical Imaging with Deep Learning}
}

@article{monai,
  title={Monai: An open-source framework for deep learning in healthcare},
  author={Cardoso, M Jorge and Li, Wenqi and Brown, Richard and Ma, Nic and Kerfoot, Eric and Wang, Yiheng and Murrey, Benjamin and Myronenko, Andriy and Zhao, Can and Yang, Dong and others},
  journal={arXiv preprint arXiv:2211.02701},
  year={2022}
}

@inproceedings{shinr,
  title={Implicit Neural Representation for Spherical Harmonics Reconstruction of Motion-Corrupted Fetal Diffusion MRI},
  author={Wu, Wenxuan and Grigorescu, Irina and Qu, Ruowen and Hajnal, Joseph V and Tournier, J-Donald and Deprez, Maria},
  booktitle={European Conference on Computer Vision},
  pages={56--72},
  year={2026},
  organization={Springer}
}

@inproceedings{scaffold,
  title={Scaffold-gs: Structured 3d gaussians for view-adaptive rendering},
  author={Lu, Tao and Yu, Mulin and Xu, Linning and Xiangli, Yuanbo and Wang, Limin and Lin, Dahua and Dai, Bo},
  booktitle={2024 IEEE/CVF Conference on Computer Vision and Pattern Recognition (CVPR)},
  pages={20654--20664},
  year={2024},
  organization={IEEE}
}

@article{dmri-shfit,
  title={Characterization of anisotropy in high angular resolution diffusion-weighted MRI},
  author={Frank, Lawrence R},
  journal={Magnetic Resonance in Medicine: An Official Journal of the International Society for Magnetic Resonance in Medicine},
  volume={47},
  number={6},
  pages={1083--1099},
  year={2002},
  publisher={Wiley Online Library}
}

@article{meta-inr,
  title={Meta-learning slice-to-volume reconstruction in fetal brain mri using implicit neural representations},
  author={Dannecker, Maik and Sanchez, Thomas and Cuadra, Meritxell Bach and Turgut, {\"O}zg{\"u}n and Price, Anthony N and Cordero-Grande, Lucilio and Kyriakopoulou, Vanessa and Hajnal, Joseph V and Rueckert, Daniel},
  journal={arXiv preprint arXiv:2505.09565},
  year={2025}
}

@article{jo1,
  title={MRI of moving subjects using multislice snapshot images with volume reconstruction (SVR): application to fetal, neonatal, and adult brain studies},
  author={Jiang, Shuzhou and Xue, Hui and Glover, Alan and Rutherford, Mary and Rueckert, Daniel and Hajnal, Joseph V},
  journal={IEEE transactions on medical imaging},
  volume={26},
  number={7},
  pages={967--980},
  year={2007},
  publisher={IEEE}
}

@article{jo2,
  title={Diffusion tensor imaging (DTI) of the brain in moving subjects: application to in-utero fetal and ex-utero studies},
  author={Jiang, Shuzhou and Xue, Hui and Counsell, Serena and Anjari, Mustafa and Allsop, Joanna and Rutherford, Mary and Rueckert, Daniel and Hajnal, Joseph V},
  journal={Magnetic Resonance in Medicine: An Official Journal of the International Society for Magnetic Resonance in Medicine},
  volume={62},
  number={3},
  pages={645--655},
  year={2009},
  publisher={Wiley Online Library}
}
}

\end{document}